\documentclass[12pt]{article}                              %% LaTeX2e

\usepackage{amsmath,amsthm,amsfonts,amscd,eucal}
\newlength{\dinwidth}
\newlength{\dinmargin}
\newsymbol\rest 1316         %% restriction symbol
\numberwithin{equation}{section}

\def\cA{{\cal A}}

\def\cD{{\cal D}}

\def\cH{{\cal H}}

\def\cO{{\cal O}}
\def\Oo{{\cal O}}

\def\cS{{\cal S}}
\def\Ss{{\cal S}}

\def\cU{{\cal U}}
\def\cV{{\cal V}}
\def\cW{{\cal W}}
\def\cX{{\cal X}}

\def\bC{{\mathbb C}}

\def\bI{{\mathbb I}}

\def\bN{{\mathbb N}}

\def\bR{{\mathbb R}}

\def\a{\alpha}
\def\b{\beta}
\def\g{\gamma}        \def\G{\Gamma}
\def\d{\delta}        \def\D{\Delta}

\def\k{\kappa}
\def\l{\lambda}       
\def\la{\lambda}
\def\m{\mu}
\def\n{\nu}

\def\p{\pi}
\def\r{\rho}

\def\t{\tau}

\def\o{\omega}        \def\O{\Omega}

\def\imply{\Rightarrow}

\newtheorem{Thm}{Theorem}[section]
\newtheorem{Cor}[Thm]{Corollary}
\newtheorem{Prop}[Thm]{Proposition}
\newtheorem{Lemma}[Thm]{Lemma}
\theoremstyle{definition}
\newtheorem{Dfn}[Thm]{Definition}

\theoremstyle{remark}

\begin{document}
%%%%%%%%%%%%%%%%%%%%%%%%%%%%%%%%%%%%%%%%%%%%%%%%%%%%%%%%%%%%%
\title{{\large \bf Wavefront Sets in Algebraic Quantum Field Theory}}
\author{ 
 {\sc  Rainer Verch}
\\[14pt]
{\normalsize Institut f\"ur Theoretische Physik,}\\
{\normalsize Universit\"at G\"ottingen,}\\
{\normalsize Bunsenstr. 9,}\\
{\normalsize  D-37073 G\"ottingen, Germany}\\
{\normalsize e-mail: verch$@$theorie.physik.uni-goettingen.de}}
\date{}
\maketitle
%% 
%%%%%%%%%%%%%%%%%%%%%%%%%%%%%%%%%%%%%%%%%%%%%%%%%%%%%%%%%%%%%%%%%%%%%%%%
\newcommand{\Cin}{C^{\infty}}
\newcommand{\Coin}{C_{0}^{\infty}}
\newcommand{\hh}{{\sf h}}
\newcommand{\rr}{{\sf r}}
\newcommand{\lcrc}{ \mbox{\footnotesize $\circ$}}
\newcommand{\nin}{\in \!\!\!\!\!/\  }          % not element of
\newcommand{\ax}{\alpha_x}
\newcommand{\axu}{\underline{\alpha}_x}        % lifted translations
\newcommand{\ayu}{\underline{\alpha}_y}
\newcommand{\afu}{\underline{\alpha}_f}    
\newcommand{\alx}{\alpha_{\lambda x}}
\newcommand{\axR}{\{\alpha_x\}_{x \in \bR^d}}  % translation group
\newcommand{\cAu}{\underline{\cal A}}          % scaling algebra
\newcommand{\Au}{\underline{A}}                % scaling alg.function
\newcommand{\Bu}{\underline{B}}
\newcommand{\Aul}{\underline{A}_{\lambda}}     % value at parameter
                                               % lambda
\newcommand{\Bul}{\underline{B}_{\lambda}}
\newcommand{\olu}{\underline{\omega}_{\lambda}}  % lifted states
\newcommand{\ooi}{\omega_{0,\iota}}            % scaling limit state
\newcommand{\Aoi}{{\cal A}_{0,\iota}}           % scaling limit algebra
\newcommand{\Hoi}{{\cal H}_{0,\iota}}          % scaling limit GNS
\newcommand{\Ooi}{\Omega_{0,\iota}}
\newcommand{\poi}{\pi_{0,\iota}}
\newcommand{\ayoi}{\alpha_y^{(0,\iota)}}       % transl. in scaling limit
\newcommand{\Rl}{{\mathbb R}}                  % real nos. (again)
\newcommand{\Nl}{{\mathbb N}}                  % integers (again)
\newcommand{\Cl}{{\mathbb C}}                  % complex nos. (again) 
%
% n-tupels of vectors and boldmath symbols
%   
\newcommand{\xx}{{\mbox{\boldmath $x$}}}
\newcommand{\xs}{{\mbox{\scriptsize\boldmath $x$}}}
\newcommand{\kk}{{\mbox{\boldmath $k$}}}
\newcommand{\kks}{{\mbox{\scriptsize\boldmath $k$}}}
\newcommand{\ks}{{\mbox{\boldmath $\xi$}}}
\newcommand{\yy}{{\mbox{\boldmath $y$}}}
\newcommand{\zz}{{\mbox{\boldmath $z$}}}
\newcommand{\ys}{{\mbox{\scriptsize\boldmath $y$}}}
\newcommand{\kp}{{\mbox{\boldmath $k'$}}}
\newcommand{\rb}{{\mbox{\boldmath $)\!\!)$}}}
\newcommand{\lb}{{\mbox{\boldmath $(\!\!($}}}
\newcommand{\zul}{\underline{{\rm z}}}
\newcommand{\kul}{\underline{{\rm k}}}
\newcommand{\ksul}{\underline{\xi}}
%
% n-tupel phase function
%
\newcommand{\eilky}{{\rm e}^{-i \lambda^{-1} {\mbox{\scriptsize\boldmath
        $k\cdot y$}}}}
%
% \renewcommand{sectionmark}[]
%
%%%%%%%%%%%%%%%%%%%%%%%%%%%%%%%%%%%%%%%%%%%%%%%%%%%%%%%%%%%%%%%%%%%%%%%%%%
%%%%%%%%%%%%%%%%%%%%%%%%%%%%%%%%%%%%%%%%%%%%%%%%%%%%%%%%%%%%%%%%%%%%%%%%%%
%%%%%%%%%%%%%%%%%%%%%%%%%%%%%%%%%%%%%%%%%%%%%%%%%%%%%%%%%%%%%%%%%%%%%%%%%%
%%%%%%%%%%%
%%%%%%%%%%%%%%%%%%%%%%%%%%%%%%%%%%%%%%%%%%%%%%%%%%%%%%%%%%%%%%%%%%%%%%%
%%%%%%%%%%%%%%%%%%%%%%%%%%%%%%%%%%%%%%%%%%%%%%%%%%%%%%%%%%%%%%%%%%%%%%%%%
\noindent
${}$\\[10pt]
\noindent
{\small  {\bf Abstract:} The investigation of wavefront sets of
  $n$-point distributions in quantum field theory has recently
  acquired some attention stimulated by results obtained with the help
  of concepts from microlocal analysis in quantum field theory in curved
  spacetime. In the present paper, the notion of wavefront set of a
  distribution is generalized so as to be applicable to states and
  linear functionals on nets of operator algebras carrying a covariant
  action of the translation group in arbitrary dimension. In the case
  where one is given a quantum field theory in the operator algebraic
  framework, this generalized notion of wavefront set, called
  ``asymptotic correlation spectrum'', is further investigated and
  several of its properties for physical states are derived. We also
  investigate the connection between the asymptotic correlation
  spectrum of a physical state and the wavefront sets of the
  corresponding Wightman distributions if there is a Wightman field
  affiliated to the local operator algebras. Finally we present a new
  result (generalizing known facts) which shows that certain spacetime
  points must be contained in the singular supports of the $2n$-point
  distributions of a non-trivial Wightman field. 
 } 
%%%%%%%%%%%%%%%%%%%%%%%%%%%%%%%%%%%%%%%%%%%%%%%
\section{Introduction}
%%%%%%%%%%%%%%%%%%%%%%%%%%%%%%%%%%%%%%%%%%%%%%%
 The wavefront set of a distribution (see the next section for a
 definition) is a mathematical
concept which proved very useful in the analysis of  partial
differential equations and, more generally, pseudo-differential operators
(see, e.g., \cite{HorFI,Dui,Hor1,Hor3,Tay}). The utility of this
concept was also realized  
 by quantum field theorists and, in fact,
early forms of this notion can be traced in literature on quantum field
 theory (see e.g.\ \cite{IaSt,Iag} and references cited therein)
 while the mathematical definition of the
 $C^{\infty}$-wavefront set in the form as it
 is used nowadays in mathematics apparently is  due to H\"ormander
\cite{HorFI}. Also, the notion of analytic
wavefront set was, parallely to its introduction in mathematics
\cite{Sat}
\footnote{We recommend that the reader consults the Notes to
  Chapter IX in \cite{HorFI} for a brief but informative review of
  the development in this branch of mathematics together with a list
  of the relevant references.},
developed in quantum field theory \cite{BrIa,Iag} and used there mainly for
the study  of analytic properties of scattering amplitudes and the
behaviour of Wightman distributions in momentum space. However, until
recently, microlocal analytic methods were apparently seldom 
 used for the study of  Wightman distributions in configuration space.
 The main reason seems to be that in the Wightman
framework of quantum field theory in Minkowski spacetime one has the
global Fourier-transform at one's disposal so that the need for a
``local'' version of Fourier-transform methods, as provided e.g.\ by
the wavefront set-concept, does not automatically arise, at least not when
considering  theories in vacuum representation.

The situation is, of course, much different when one wishes to study
quantum field theory in curved spacetime (see \cite{Wald2,Ful} as
general references), where one is faced with all the difficulties
brought about by the absence of translational symmetries, and in the
generic case, the absence of any spacetime symmetry. As is well-known,
the notions of a vacuum state, and of a particle, are in quantum field
theory in Minkowski spacetime tied to invariance and spectrum
condition with respect to the translation group (see e.g.\
\cite{SW,HK,Haag,Bor2}).  The spectrum
condition is of particular importance since it expresses dynamical
stability of quantum field theories. In quantum field theory in
Minkowski spacetime, the spectrum condition can be formulated by means
of globally Fourier-transforming the action of the translations on the
observables and by suitably restricting the resulting
Fourier-spectrum of vacuum expectation values. But in curved
spacetimes, this is not possible, and one has to look for other ways
of formulating conditions of dynamical stability. The wavefront set
describes  ultra-local remnants of the singular contributions of a
distribution in Fourier-space, and this property allows to give 
appropriate local versions of restrictions on the Fourier-spectrum for
dynamically stable states also in quantum field theory in curved
spacetime.
But even though this line of thought was to some extend developed in
the mathematical
literature \cite{HorFI,DH}, it has  only quite
recently been investigated seriously within mathematical physics, beginning
with Radzikowski's result which says  that the two-point function
of a state of the free scalar field on a curved (globally hyperbolic)
spacetime is of Hadamard form exactly if its wavefront set has a
structure which is formally the same as that displayed by the
wavefront set of the free field vacuum in Minkowski spacetime
\cite{Rad1}. This result bears some importance since several
works provide, from various directions, evidence that Hadamard states of
the free field in curved spacetime are to be viewed as physical,
dynamically stable states (see \cite{Wald2,Ful} and references cited
there, and also \cite{Jun,Ver}).

 Radzikowski then proposed that
generally the spectrum condition for Wightman fields in Minkowski
spacetime should in curved spacetime be replaced by restrictions on
the wavefront sets of the $n$-point Wightman functions. His first
proposal for such restrictions (called WFSSC, ``wavefront set spectrum
condition'') underwent some changes \cite{Koh,Rad1,BFK}; in their current
form, they are called $\mu$SC, ``microlocal spectrum condition''
\cite{BFK}. Subsequently, the use of wavefront set methods and the
closely related pseudo-differential operator techniques led to a few
interesting results in quantum field theory in curved
spacetime. Examples are Radzikowski's local-to-global singularity
theorem for two-point functions of the free scalar field \cite{Rad2},
Junker's proof that adiabatic vacuum states of the Klein-Gordon field
are Hadamard states \cite{Jun}, the covariant definition of
Wick-products of the scalar field \cite{BFK} and the program by
Brunetti and Fredenhagen which develops the Epstein-Glaser framework
for the perturbative construction of interacting quantum field
theories and establishes renormalizability of $\phi^4_4$ on curved
spacetime \cite{BF}; finally, there are results concerning the
stability of quantum fields in spacetimes which contain compactly
generated Cauchy-horizons \cite{KRW} (such spacetimes have been
proposed as modelling situations where ``time-machines are set into
operation'' \cite{Haw}). 

These developments suggest to further pursue and utilize wavefront set
concepts and techniques in general quantum field theory (both in
curved and flat spacetimes). In its present formulation, the notion of
the wavefront set applies to distributions and hence to theories which
are formulated in terms of pointlike fields and their corresponding
Wightman distributions. From the point of view of general local
quantum field theory
\cite{HK,Haag} the description of a theory in terms of pointlike
fields is, however, not completly intrinsic, and thus a notion of
wavefront set which depends on the use of pointlike fields
isn't a completely intrinsic concept in general quantum field theory
(in operator algebraic formulation) either. We therefore attempt to
generalize the notion of wavefront set in such a way that it becomes
an intrinsic concept in algebraic quantum field theory, like the
spectrum of the translation group of a quantum field theory in
Minkowski spacetime. In the present work, we shall in this attempt
concentrate on algebraic quantum field theory on Minkowski
spacetime. The methods we use here can in principle be generalized to
quantum field theories in curved spacetime. In fact, the future
application to quantum 
field theory in curved spacetime is the main motivation for the present study
which ought to be seen as a first step in a program which eventually
aims at establishing structural results of quantum field theory (e.g.\
spin-statistics theorems) in curved spacetime 
with the help of microlocal analytic methods. 

The starting point of our work will be the observation (cf.\ Prop.\ 2.1)
that the wavefront set of a distribution can be characterized in a
novel way which emphasizes its role as an asymptotic notion of
(Fourier-) spectrum for the action of the translation
group. The idea is, roughly, the following. Suppose we have a dual
pair $L',L$ of vector spaces and a representation $\bR ^n\owns x \mapsto
\t_x$ of the additive group $\bR^n$ by automorphisms of $L$. Assume
further that the functions $x \mapsto u(\t_x(f))$, $u \in L'$, $f \in
L$, are continuous. The spectrum ${\rm sp}^{\t}u$ of an element $u \in
L'$ with respect to the action $\t_x$ may then be defined as the
support of the Fourier-transform of $x \mapsto u \lcrc \t_x$, i.e.\ as
the closed union of the supports (in the sense of distributions) of
the Fourier-transforms of all functions $x \mapsto u(\t_x(f))$, $f \in
L$. This means that $\xi \in \bR^n$ is not contained in the spectrum
of $u$ if one can find some neighbourhood $V$ of $\xi$ in $\bR^n$ so
that
\begin{equation}
\lim_{\l \to 0}\,\int {\rm e}^{-ik\cdot k}h(\l x)u(\t_x(f))\,d^nx = 0
\end{equation}
holds for all $k \in V$ and all $f$ in $L$, where $h \in \cD(\bR^n)$
is a test-function with $h(0) = 1$. 

It is obvious that in this manner spectral properties of the action of
$\t_x$ on $u$ are tested globally. To illustrate how one may test the
Fourier-spectrum behaviour when $u$ is ``asymptotically localized at a
point'', we specialize the setting for a moment and take $L$ to be the
space of test-functions $\cD(\bR^n)$ and $L' =\cD'(\bR^n)$ with the
usual action of the translations $(\t_x f)(x') = f(x' -x)$, $f \in
\cD(\bR^n)$. On $\cD(\bR^n)$ there act also the dilations $(\d_{\l}
f)(x') = \l^{-n}f(\l^{-1}x')$, $\l > 0$. When we act with the induced
action $\d_{\l}'u = u \lcrc \d_{\l}$ on $u$, then $\d_{\l}'u$ becomes
asymptotically concentrated at the origin as $\l$ approaches
$0$. However, in general the limit of $\d_{\l}'u$ for $\l \to 0$ will
not exist. Nevertheless, we may replace $u$ by $\d_{\l}'u$ in
(1.1). That means, instead of looking at the limiting behaviour as $\l
\to 0$ of $\int {\rm e}^{-ik\cdot x}h(\l x)u(\t_x(f))\,d^nx$, we
investigate the asymptotic behaviour of the expression 
\begin{equation}
 \int {\rm e}^{-ik\cdot x}h(\l x)(\d_{\l}'u)(\t_x(f))\,d^nx
 = \l^{-n}\int {\rm e}^{-i\l^{-1}k\cdot
   x}h(x)u(\t_x(\d_{\l}f))\,d^nx\,, \quad f \in \cD(\bR^d)\,,
\end{equation}
as $\l \to 0$. It turns out that the point $(0,\xi)$ is in the
complement of the wavefront set of $u$ exactly if there are a
neighbourhood $V$ of $\xi$ and an $h \in \cD(\bR^n)$ such that for all
$k \in V$ the expression in (1.2) approaches 0 faster than $O(\l^N)$
as $\l \to 0$ for any $N \in \bN$, cf.\ Prop.\ 2.1. This allows it to
interpret the wavefront set of $u$ as an asymptotic form of the
spectrum of $u$ with respect to translations when $u$ is
asymptotically localized at a point. (This point was here the origin,
but that may be changed simply via replacing $\d_{\l}$ by
$\t_{x'}\d_{\l}$ so that $u \lcrc \t_{x'}\d_{\l}$ becomes localized at
$x'$ for $\l \to 0$.) 
 
Thus it is already apparent that the concept of wavefront set may be
generalized from distributions to elements $u$
in the dual space $L'$ of a vector space $L$ on which translations and
dilations act (in a suitably continuous manner); it could even be
generalized to more general group actions and a microlocal analysis of
general automorphism groups could be developed along this line.
 But we would now like to
indicate that the notion of localization is the central one for our
concern, the generalization of the wavefront set concept to the
setting of algebraic quantum field theory. 

The basic structure of a quantum field theory in the operator
algebraic setting is that of an inclusion-preserving map $\cO \to \cA(\cO)$
assigning to each $\cO \subset \bR^d$ of a $d$-dimensional spacetime
($d \ge 2$) a $C^*$-algebra containing the observables which are
localized in the spacetime region $\cO$, i.e.\ which can be measured
during times and locations within $\cO$ \cite{HK,Haag}. Such a map is
called a {\it net of local algebras}. In quantum field theory in flat Minkowski
spacetime we additionally assume that the translations act by
automorphisms on the local algebras. That means there is a
representation of the additive group $\bR^d$ by automorphisms $\a_x$,
$x \in \bR^d$, on the $C^*$-algebra $\cA = \cA(\bR^d)$ generated by all
the local algebras $\cA(\cO)$ which acts covariantly,
\begin{equation}
   \a_x(\cA(\cO)) = \cA(\cO + x)\,,
\end{equation}
and is suitably continuous. By way of comparison we note that the
space $\cD(\bR^d)$ is also generated by a net of local test-function
spaces, $\cO \to \cD(\cO)$, and the translations act covariantly:
$\t_x(\cD(\cO)) = \cD(\cO + x)$. On the local test-function spaces we
also have the covariant action of the dilations, $\d_{\l}(\cD(\cO)) =
\cD(\l\cO)$.

 However, although there is a special class of quantum
field theories 
admitting also  covariant automorphic actions 
of the dilations on their nets of local
algebras, a general quantum field theory describing e.g.\ massive
elementary particles cannot be expected to be in this way
dilation-covariant.
 This appears at first sight to be an obstruction to
the attempt of generalizing the asymptotic localization of a
distribution at a point to linear functionals and hence, states, on
the algebra $\cA$. But in \cite{BV1} a method for analyzing the short
distance behaviour of states (positive linear functionals)
on $\cA$ was developed which circumvents the problem that one lacks a
canonical notion of dilations as actions on a generic net of local
algebras.
(We will not give a systematic account of this ``scaling algebra''
method in this paper, yet the basic elements of this approach appear
in Sec.\ 4  where our
results make contact with  material in \cite{BV1}, so the present
work is self-contained. The interested reader is referred to the
references \cite{BV1,BV2,Bu1,Bu2,Bu3} for further discussion and
results of the ``scaling algebra'' framework.) Roughly, the idea is to
collect all functions $\l \to A_{\l} \in \cA$ depending on a positive
real scaling parameter $\l$ which are uniformly bounded and have the same
localization properties as elements of the local algebras $\cA(\cO)$
would have under a covariant action of the dilations, that is, $A_{\l}
\in \cA(\l \cO)$, $\l > 0$, for some arbitrary bounded region
$\cO$. Such functions will in Sec.\ 3 be denoted as families
$(A_{\l})_{\l>0}$ and referred to as ``testing-families''.
 The counterpart at the level of test-functions is to consider
not only functions of the positive reals into $\cD(\bR^d)$ which are of
the form $\l \mapsto \d_{\l}f$, $\l > 0$, for any $f \in \cD(\bR^d)$,
but any suitably bounded family $(f_{\l})_{\l > 0}$ with $f_{\l} \in
\cD(\l \cO)$ for some bounded region $\cO$. The main content of Prop.\
2.1 is that, when taking in (1.2) all such families $(f_{\l})_{\l >
  0}$ in place of $\d_{\l}f$, $\l > 0$, one finds that the described
criterion for $(0,\xi)$ to be in the complement of the wavefront set
of $u$ remains valid. This observation motivates our definition in Sec.\ 3 of
the ``asymptotic correlation spectrum'' of a continuous linear
functional $\varphi$ on $\cA$ as a natural generalization of the
concept of the wavefront set of a distribution, and as an asymptotic
version of the Fourier-spectrum of $\varphi$ with respect to the
action of the translations in the case where the functional $\varphi$
is asymptotically localized at (simultaneously) several points in $\bR^d$.

We should like to point out that the idea of characterizing the
wavefront set of a distribution with the help of testing families
$(f_{\l})_{\l > 0}$ in a way similar to Prop.\ 2.1 is not entirely new,
it appears e.g.\ in the description of the ``asymptotic frequency
set'' in \cite{GuiSt}. However, our approach is novel in emphasizing
the ``asymptotic spectrum'' point of view, allowing immediate
generalization to functionals and group actions on vector spaces.
Moreover we remark that the asymptotic correlation spectrum tests
spectral properties of the states of a given quantum field theory and
not directly those of  the corresponding 
``scaling limit states'' and
``scaling limit theories'' in the sense of \cite{BV1} although there
is a relation, as we discuss in Section 4.

This work is organized as follows. Section 2 establishes the results
already mentioned concerning the description of the wavefront set of
distributions. In Section 3 we introduce, motivated as indicated 
by the results
of Proposition 2.1, the notion the asymptotic correlation spectrum of
a continuous linear functional on $\cA$.  Section 4 is concerned
 with a study of asymptotic
correlation spectra in the setting of quantum field theories in
Minkowski spacetime fulfilling locality and spectrum condition. The
latter two  properties are found to imply ``upper bounds'' for the
asymptotic correlation spectra of physical
states. Moreover, constraints on the asymptotic correlation spectra of
physical states are shown to imply certain properties of the
corresponding ``scaling limit states'' in the sense of \cite{BV1}. In
Sec.\ 5 we assume that there is a Wightman field affiliated to a net
of local (von Neumann) algebras, and we compare the asymptotic
correlation spectrum of a physical state with the wavefront sets of
its associated Wightman distributions. It is shown that the wavefront
sets of the Wightman functions provide ``lower bounds'' for the
asymptotic correlation spectra. We also show that if the Wightman
field is non-trivial, i.e.\ the field operators are not just multiples
of the unit operator, then for each $n\in \bN$
the essential support of the $2n$-point distribution associated with
any separating state  vector in the field domain must contain points
of a certain type, and thus has non-empty wavefront set.
The article is concluded by summary and outlook in the final Section 6.

%%%%%%%%%%%%%%%%%%%%%%%%%%%%%%%%%%%%%%%%%%%%%%%%%%
\section{On the wavefront set of distributions}
\setcounter{equation}{0}
%%%%%%%%%%%%%%%%%%%%%%%%%%%%%%%%%%%%%%%%%%%%%%%%%%
As discussed in the Introduction, we wish to introduce in the present
section a characterization of the wavefront set of a distribution which
may be viewed as an asymptotic spectrum with respect to the action of
the translation group. It bears some reminiscence to the ``frequency
set'' of a distribution introduced by Guillemin and Sternberg
\cite{GuiSt}.
\\[6pt]
{\it Notation.}\quad In the  following discussion, $m \in \Nl$
is arbitrary but kept fixed, thus we write $\cD \equiv \cD (\Rl^m)$,
$\cS \equiv \cS(\bR^m)$, $\cD' \equiv \cD'(\Rl^m)$, etc. 
 We denote by $\tau_y$,
$y \in \Rl^m$, the action of the translations on test-functions:
\begin{equation}
(\tau_y f)(x) := f(x -y) \,, \quad x,y \in \Rl^m,\ f \in \cD\,.
\end{equation}
We often write $<\!\!u,f\!\!> \ \equiv u(f)$, $f\in \cD$, $u \in \cD'$, for the
dual pairing between distributions and test-functions.

The reflection of a test-function $f$ with respect to the origin
will be denoted by
\begin{equation}
 {}^r\!\!f(x) := f(-x)\,, \quad x \in \Rl^m \,.
\end{equation}
(In the literature,  $\check{f}$ is often used to denote
the reflection of $f$.)

The Fourier-transform of a test-function $f$ is defined by
\begin{equation}
 \widehat{f}(k) := \int {\rm
 e}^{-ik\cdot x}f(x)\,d^mx\,,
\quad k \in \Rl^m\,,
\end{equation}
where $k \cdot x$ denotes the Euklidean scalar product of elements
in $\Rl^m$.

We also use the following convention which is more or less
standard.
Let $\varphi_{k}: \Rl^+ \to \Cl$ be a family of functions
parametrized by elements $k$ in some set $K$. Then the statement
that
\begin{equation}
\varphi_{k}(\la) = O^{\infty}(\la) \ \ \ {\rm as}\ \ \la \to 0 \ \ 
{\rm uniformly\ in}\ k \in K
\end{equation}
is an abbreviation of the following statement:

For each $N \in \Nl$ there exist $C_N > 0$ and $\la_N > 0$ such that
\begin{equation}
 \sup_{k \in K}\, |\varphi_{k}(\la)|\ <\ C_N \cdot \la^N \ \ \ 
{\rm for \ all}\ \  0 < \la < \la_N\,.
\end{equation}

It should be observed that if (2.4) holds, then there holds
equivalently for all $\m \in \bR$ and any $\n > 0$, $\varphi_k(\l) =
O^{\infty}(\l^{\n})$ as well as $\l^{\m}\varphi_k(\l) =
O^{\infty}(\l)$ for $\l \to 0$, uniformly in $k \in K$. It is also not
difficult to check e.g.\ the following: When we have a family of
functions $a_{\k}: \bR^+ \to \bC$ indexed by real numbers $\k$ such
that $a_{\k}(\l) = O^{\infty}(\l)$ as $\l \to 0$ for each $\k$, and if
$b : \bR^+ \to \bC$ is another function having the property that there
is some $c \in \bR$ and for
every $\n > 0$ some $\k = \k(\n)$ with $\l^{c\k}a_{\k}(\l) - b(\l) =
O(\l^{\n})$ as $\la \to 0$,  then it follows that $b(\l) =
O^{\infty}(\l)$ as $\la \to 0$. 

We shall now introduce a collection of families
 $(f_{\l})_{\l  > 0}$ 
of test-functions $f_{\l}$ indexed by a real positive 
parameter $\l$; we will call such families ``testing-families''. We
shall use the notation $\lb f_{\l} \rb \equiv (f_{\l})_{\l   > 0}$.
 Observe that thus, by convention,  
the use of parentheses in bold print means that we are considering the
whole testing family (i.e.\ a mapping from $\bR^+$ into $\cD$), as
opposed to e.g.\ writing $u(f_{\l})$, where a distribution $u$ is
 evaluated on the member of a testing family at some particular
 parameter value $\l$.
 For $x \in \Rl^n$ and $\Oo$ a bounded ,
open neighbourhood of $0 \in \Rl^m$,  we define the set
\begin{equation}
 {\bf F}_x(\Oo) := \left\{ \lb f_{\la}\rb : f_{\la} \in \cD,\ 
 {\rm supp}\, f_{\la} \subset \la \Oo + x, \ \sup_{\la}\,
 ||\,f_{\la}\,|| < \infty \right\} \,,
\end{equation}
where $||\,f\,|| := \sup_{y \in \bR^m}|f(y)|$. As our collection of
testing families we then take ${\bf F}_x := \bigcup_{\cO}{\bf F}_x(\cO)$.

We should finally note that we will usually denote variables in
configuration space by letters $x,x',y$ etc., while reserving the
letters $\xi,k,\ell$ for variables in Fourier-space.

 Let us also
recall the definition of the {\it wavefront set} $WF(u)$ of a
distribution $u \in \cD$: $WF(u)$ is the complement set in $\bR^m
\times (\bR^m \backslash \{0\})$ of all those pairs $(x,\xi) \in \bR^m
\times (\bR^m \backslash \{0\})$ having the property that there exists
some $\chi \in \cD$ with $\chi(x) \ne 0$ and an open neighbourhood $V$
of $\xi$ (in $\bR^m\backslash \{0\}$) such that there holds
\begin{equation}
 \sup_{k \in V}\, |\widehat{\chi u}(\l^{-1}k)| = O^{\infty}(\l)
 \quad {\rm as}\quad \l \to 0\,.
\end{equation}
Here, $\widehat{\chi u}$ is the Fourier-transform of the distribution
$\chi u$, which may be expressed as $\widehat{\chi u}(k) =
u(e_k\chi)$ with $e_k(y) := {\rm e}^{-ik\cdot y}$. This form of  
definition of the wavefront set can be found in \cite{Dui}. We refer
to this reference and the e.g.\ the monographs  \cite{Hor1,Hor3,Tay} 
 for considerable further discussion on the
properties of the wavefront set and its use in studying partial (or pseudo-)
differential operators.
\begin{Prop}
Let $x \in \Rl^m$, $\xi \in \Rl^m \backslash \{0\}$, and $u \in \cD'$.
Then the following statements are equivalent.
\\[6pt]
(a)\quad $(x,\xi) \nin  WF(u)$
\\[6pt]
(b)\quad There exist an open neighbourhood $V$ of $\xi$ and an $h \in \cD$
with $h(0) =1$, such that for each family
$\lb f_{\la} \rb \in {\bf F}_x$ there holds
\begin{equation}
\int {\rm e}^{-i\la^{-1}k\cdot  y}h(y)<\!\!u,\tau_y f_{\la}
\!\!>\,d^my\ =\ O^{\infty}(\la)
\ \ {\rm as}\ \ \la \to 0
\end{equation} 
uniformly in $k \in V$.
\\[6pt]
(c)\quad There exist an open neighbourhood $V$ of $\xi$, an  $h \in \cD$
with $h(0) =1$,  and some $g \in \cD$ with $\widehat{g}(0) =1$ such that
for all $p \ge 1$ it holds that
\begin{equation}
\int {\rm e}^{-i\la^{-1}k\cdot y}h(y)<\!\!u,\tau_y g^{(p)}_{\la}
\!\!>\,d^my\ =\ O^{\infty}(\la)
\ \ {\rm as}\ \  \la \to 0
\end{equation} 
uniformy in $k \in V$, where
\begin{equation}
 g^{(p)}_{\la}(x') := g(\la^{-p}(x' - x)) \,, \quad \la > 0,\,x' \in \bR^m\,.
\end{equation}  
\end{Prop}
\begin{proof}
 The first step is to facilitate the proof by
demonstrating that it is sufficient to consider the case $x=0$. To
this end, we notice that $(x,\xi) \nin WF(u)$ if and only if  
$(0,\xi) \nin WF(u \lcrc \tau_{-x})$ by the well-known transformation
properties of the wavefront set. Secondly, we notice that requiring
for $u$ the condition (2.8) to hold for all
 $\lb f_{\la}\rb \in {\bf F}_x$ is
equivalent to demanding that (2.8) holds with $u \lcrc \tau_{-x}$ in
place of $u$ for all $\lb f_{\la}\rb \in {\bf F}_0$, as can be seen
from
\begin{equation}
 <\!\! u \lcrc \tau_{-x},\tau_{y}f_{\la}\!\!>\ =\ 
 <\!\!u,\tau_y \tau_{-x}f_{\la}\!\!> 
\end{equation}
and the observation that $\lb f_{\la}\rb \mapsto \lb \tau_{-x} f_{\la}\rb$
induces a bijective map from ${\bf F}_x$ onto
 ${\bf F}_0$.
By the same type of argument one concludes that replacing in (2.9) $u$ by
$u \lcrc \tau_{-x}$ is equivalent to replacing in the definition (2.9)
 of $g^{(p)}_{\la}$ the $x$ by $0$.
Thus it suffices to prove the claimed equivalences for the case $x=0$.
We will use the notation ${\bf F} \equiv {\bf F}_0$. 
\\[6pt]
To carry on, it is convenient to collect first a few auxiliary
 results.
\begin{Lemma}
($\a$)\quad Let  $\lb f_{\la}\rb \in {\bf F}$. Then
there is $c >0$  such that
\begin{equation}
\sup_{k \in \Rl^m}|\widehat{f_{\la}}(\la^{-1}k)| \le c \cdot \la^m\,.
\end{equation}
($\b$)\quad Let $w_{\la}$, $1 >\la >0$, be a family of smooth functions on
$\Rl^m$ fulfilling the bound
\begin{equation}
 |w_{\la}(\la^{-1}k)| \le c\cdot (|k|+ 1 + \la^{-1})^q 
\end{equation}
for suitable numbers $c>0$ and $q \in \Rl$. Assume additionally
that there is an open neighbourhood $V'$ of some $\xi \in \Rl^m
\backslash \{0\}$ such that
\begin{equation}
 w_{\la}(\la^{-1}k') = O^{\infty}(\la)\ \  {\rm as}\ \ \la \to 0
\end{equation}
uniformly in $k' \in V'$. Then  for each open neighbourhood $V$ of
 $\xi$ contained in any compact subset of $V'$ and for all $\phi \in \Ss$
 one has
\begin{equation}
(\phi *  w_{\la})(\la^{-1}k) = O^{\infty}(\la)\ \  {\rm as}\ \ \la \to 0
\end{equation}
uniformly in $k \in V$.
\end{Lemma}
\begin{proof}        
 ($\a$) We have $\lb f_{\l} \rb \in {\bf F}(\cO)$ for some bounded set
 $\cO$, hence we obtain
\begin{equation}
|\widehat{f_{\la}}(\la^{-1}k)|  \le 
 \left|\, \int {\rm e}^{-i\la^{-1}k\cdot y}f_{\la}(y)\,d^ny \,\right|
 \le {\rm vol}(\l \cO) \sup_{\l}\,||\,f_{\l}\,||\,,
\end{equation}
implying the assertion.
\\[6pt]
($\b$) Let $U$ be an open, bounded neighbourhood around the origin in
$\bR^m$ and $V$ a bounded open neighbourhood of $\xi$ such that
$\overline{V} \subset V'$. Define two functions $\chi_U,\chi$ on
$\bR^m$ by $\chi_U$ = characteristic function of $-U$, $\chi = 1
-\chi_U$. After a change of variables one gets
\begin{equation}
(\phi * w_{\l})(\l^{-1}k) = \frac{1}{\l^{m/2}} \int(\chi_U(\ell) +
\chi(\ell)) \phi(\l^{-1/2}\ell)w_{\l}(\l^{-1}(k
-\l^{1/2}\ell))\,d^m\ell\,.
\end{equation}
Since $\phi \in \cS$, it holds that 
\begin{equation}
\int |\chi(\ell)\phi(\l^{-1/2}\ell)|(\l^{-1} + |\ell|)^s\,d^m\ell =
O^{\infty}(\l) \quad {\rm as} \quad \l \to 0
\end{equation}
for all $s \ge 0$. Moreover, for sufficiently small $\l$ one has $V +
\l^{1/2}U \subset V'$, thus
\begin{equation}
\sup_{\ell \in \bR^m}\,|\chi_U(\ell)w_{\l}(\l^{-1}(k - \l^{1/2}\ell))|
= O^{\infty}(\l) \quad {\rm as} \quad \l \to 0
\end{equation}
uniformly in $k \in V$. This proves the claim.
\end{proof}
We return to the proof of Proposition 1 and begin with:\\[6pt]
(a) $\Rightarrow$ (b). 
Let $(0,\xi) \nin WF(u)$. Then there exist an open neighbourhood $V'$
of $\xi$ and a function $\chi \in \cD$ which is equal to 1 in an open
neighbourhood $U$ of 0, with the property that
\begin{equation}
 \widehat{\chi u}(\la^{-1}k') = O^{\infty}(\la)\ \ {\rm as}\ \ \la
 \to 0
\end{equation}
uniformly in $k' \in V'$. Now let $h \in \cD$ with $h(0) =1$ and a convex
open neighbourhood $\Oo$ of 0 be chosen so that ${\rm supp}\, h + \Oo
\subset U$. Then we find
that the support properties of $h$ and  of members $\lb f_{\l} \rb$ of
 ${\bf F}(\Oo)$ imply for all $k \in \Rl$ and $1 > \la >0$,
\begin{eqnarray}
 \int \,{\rm e}^{-i \la^{-1}k\cdot y}h(y)<\!\!u,\tau_y f_{\la}\!\!>
 d^my    
 & = & \int h(y)\,{\rm e}^{-i \la^{-1}k\cdot y}<\!\!\chi u,\tau_y
 f_{\la}\!\!> d^my
\\
& = & \widehat{h} * (\widehat{{}^r\!\!f_{\la}}\cdot \widehat{\chi
 u})(\la^{-1}k)\,.
\nonumber
\end{eqnarray}
Now observe that $w_{\la} := \widehat{{}^r\!\!f_{\la}}\cdot \widehat{\chi
 u}$ fulfills by Lemma 2$(\a)$ and due
 to the fact that $\widehat{\chi u}$ is polynomially bounded the
 assumptions of Lemma 2($\b$). Whence we may apply Lemma 2($\b$) to the
 effect that the last expression in (2.21) is of order $O^{\infty}(\la)$
uniformly for $k$ in some open neighbourhood $V$ of $\xi$ as $\la \to
 0$. 
\\[6pt]
(b) $\Rightarrow$ (c).
Assume that condition (b) holds for suitable choices of $V$, $h$
  and $\Oo$. Let $g \in \cD$ have ${\rm supp}\,g \subset \cO$. Then for
 each $p \ge 1$ the testing-family $\lb f_{\l} \rb$ defined by
\begin{equation}
 f_{\l} := g^{(p)}_{\l}\,, \quad \l > 0\,,
\end{equation}
is contained in ${\bf F}(\cO)$, and this implies (c).
\\[6pt]
(c) $\Rightarrow$ (a).
We now assume that (c) holds with suitable choices of $V$, $h$ and
$g$. Then we choose some $\chi \in \cD$ which is equal to 1 on a
ball centered around the origin containing
 the set ${\rm supp}\,h + {\rm supp}\,g$. Note also that
$\widehat{g}(0) = 1$ implies $\widehat{{}^r\!\!g}(0) = 1$.  
Since for some $c> 0$ and $q \in \bR$ it holds that $|\widehat{\chi
  u}(k)| \le c (|k| + 1)^q $, we
obtain with the help of the mean value theorem and suitable constants
$c_1,c_2 > 0$,
\begin{eqnarray} & &
 \frac{1}{\l^{pm}}
\int {\rm e}^{-i \la^{-1}k\cdot y}h(y)<\!\!u,\tau_yg_{\la}^{(p)}
\!\!>\, d^my - \widehat{hu}(\la^{-1}k)\nonumber  \\
& = & \int \widehat{h}(\ell)\left( \widehat{{}^r\!\!g}(\la^p(\la^{-1}k -
  \ell)) -1 \right) \cdot \widehat{\chi u}(\la^{-1}k - \ell)\,
d^m\ell
\nonumber
\\
& \le & c_1\,\int
|\widehat{h}(\ell)|\,\l^{p -1}\,|k - \l \ell|\cdot(\l^{-1}|k| + |\ell|
 + 1 )^q \,d^m\ell\nonumber \\
 & \le & c_2\,\l^{p - 1 - q}\int |\widehat{h}(\ell)|
(1 + |\ell|)^{q + 1}\,d^m\ell \nonumber\\
& = & O(\la^{p -(q+1)}) \quad {\rm as} \quad \la \to 0 
\end{eqnarray}
uniformly for $k$ in any fixed bounded subset of $\Rl^m$.
In view of our assumption that (c) holds,
 this last estimate implies that there is some bounded
neighbourhood $V_1 \subset V$ of $\xi$ such that for
arbitrary $p \ge 1$,
\begin{equation}
 \widehat{hu}(\la^{-1}k) = O(\la^{p - (q+1)}) \quad {\rm as} \quad
 \la \to 0
\end{equation}
holds uniformly in $k \in V_1$. But since $p \ge 1$ is arbitrary and $q
\in \Rl$ is fixed, this means that $(0,\xi) \nin WF(u)$.
This completes the proof. 
\end{proof}
The fact that in (2.8) the Fourier-transform of $y \mapsto \,
<\!\!u, \tau_yf_{\la}\!\!>$ is ``windowed'' by the function $h$
ensures that only local properties of $u$ near the point $x$ are
tested, and so the role of $h$ is to localize $u$ near $x$. Thus one
expects that the behaviour (2.8) and (2.9) of $u$ is not changed if $h$ is
replaced by $\phi \cdot h$ when $\phi$ is any element in
$C^{\infty}(\bR^m)$. This turns out to be indeed the case.

\begin{Prop}  
Let $x \in \Rl^m$, $\xi \in \Rl^m\backslash \{0\}$ and $u \in \cD'$.
Assume that condition (b) of Proposition 1 is fulfilled with suitable
choices of $V$ and $h$. Then condition (b) of
Proposition 1 holds also if $h$ is replaced by $\phi\cdot h$ for any
$\phi \in C^{\infty}(\bR^d)$, and if at the same time $V$ is
 replaced by any open
neighbourhood $V_1$ of $\xi$ such that $\overline{V_1}$ is compact and
contained in $V$.

The analogous statement holds also for condition (c) of Proposition 2.1.
\end{Prop}
\begin{proof}
 Again it is sufficient to consider the case $x =0$. Let
 $\lb f_{\la} \rb \in {\bf F}(\cO)$ and 
let $\chi \in \cD$ be equal to 1 on
 a neighbourhood of the closure of $\Oo + {\rm supp}\,h$. Then it is
 easy to check that the family of smooth functions $w_{\la}$, $1 > \la
 > 0$, defined by
\begin{equation}
w_{\la} := \widehat{h} * (\widehat{{}^r\!\!f_{\la}} \cdot
\widehat{\chi u})
\end{equation}
fulfills the assumptions of Lemma 2($\b$).
Moreover, $\chi \phi \in \cD$, and we have, using Lemma 2($\b$),
 for each open, relatively
compact neighbourhood $V_1$ of $\xi$ with $\overline{V_1} \subset V$ that
\begin{eqnarray}
& & \hspace*{-1.5cm}\int
 {\rm e}^{-i \la^{-1}k\cdot y}(\phi \cdot h)(y)<\!\! u,\tau_y
 f_{\la}\!\!> d^my \\
 & = & \int
 {\rm e}^{-i \la^{-1}k\cdot y}(\chi\phi \cdot h)(y)<\!\!\chi u,\tau_y
 f_{\la}\!\!> d^my \nonumber \\
 & = & \widehat{\chi\phi}*w_{\la}(\la^{-1}k') \  = \  O^{\infty}(\la)
 \quad {\rm as} \quad \la \to 0  \nonumber
\end{eqnarray}
uniformly in $k \in V_1$.
The argument for the case of condition (c) of Proposition 1 is analogous.
\end{proof}

%%%%%%%%%%%%%%%%%%%%%%%%%%%%%%%%%%%%%%%%%%%%%%%%%%
\section{The asymptotic correlation spectrum }
%%%%%%%%%%%%%%%%%%%%%%%%%%%%%%%%%%%%%%%%%%%%%%%%%%
\setcounter{equation}{0}
%%%%%%%%%%%%%%%%%%%%%%%%%%%%%%%%%%%%%%%%%%%%%%%%%% 
We shall in this section present our definition of the ``asymptotic
correlation spectrum'' which is a generalization of the wavefront set in
algebraic quantum field theory. To this end, we must first of all
describe the algebraic quantum field theory which we
are going to consider, namely, a translation covariant theory
on $d$-dimensional Minkowski-spacetime where $d \ge 2$. More
precisely, we assume that we are given a net $\cO \to \cA(\cO)$ of
$C^*$-algebras indexed by the double cone regions $\cO \subset
\bR^d$.\footnote{A double cone $\cO$ is a set of the form $\cO = (V_+
 + x) \cap (-V_+ + y)$ and $y \in V_+$ where $V_+ = \{
  (x^0,x^1,\ldots,x^{d-1}) \in \bR^d
: (x^0)^2 - \sum_{j=1}^{n-1}(x^j)^2 > 0,\ x^0 > 0\}$ is
  the open forward lightcone; $\overline{V}_+$ is its closure.}
 Thus the map $\cO \to \cA(\cO)$ assigns to each double cone
$\cO$ a $C^*$-algebra such that the condition of isotony holds,
\begin{equation}
 \cO_1 \subset \cO_2 \imply \cA(\cO_1) \subset \cA(\cO_2)\,.
\end{equation}
It will be recalled that this is motivated by the idea to view
$\cA(\cO)$ as the algebra generated by the observables which can be
measured at times and locations in the spacetime region $\cO$, cf.\
\cite{HK,Haag} for further discussion.
 
Moreover, we make the assumption that there is a representation $\bR^d
\owns x \mapsto \ax$ of the translation group acting by automorphisms on
$\cA := \overline{\bigcup_{\cO}\cA(\cO)}^{C^*}$, the so-called
quasilocal algebra of observables. This representation is required to
act covariantly,
\begin{equation}
\ax(\cA(\cO)) = \cA(\cO + x)
\end{equation}
for all $x \in \bR^d$ and all double cones $\cO$. A further assumption
which we add here is that the group action is strongly continuous,
meaning that
\begin{equation}
||\,\ax(A) - A \,|| \to 0 \quad {\rm for} \quad x \to 0
\end{equation}
holds for all $A \in \cA$. (In mathematical terms, $(\cA,\{\ax\}_{x
  \in \bR^d})$ is a $C^*$-dynamical system.) As will be pointed out
  later, this strong continuity requirement is not really necessary and could
  in more special situations be replaced by weaker versions. However,
  when starting with the $C^*$-algebraic setting, it is a natural assumption.  

Other assumptions which are standard in quantum field theory like
locality and existence of a vacuum state (see \cite{HK,Haag}) are not
needed for the moment but will be introduced later.

With the given theory $(\cO \to \cA(\cO),\axR)$ we can now associate
testing-families  $(A_{\l})_{\l > 0}$
 in the following way: We define for each double cone
region $\cO$ in $\bR^d$ and $x \in \bR^d$ the set
\begin{equation}
{\bf A}_x(\cO) := \left\{( A_{\l})_{\l > 0} : A_{\l} \in \cA(\l \cO + x), \ 
      \sup_{\l>0}\,||\, A_{\l} \,|| < \infty ,\ A_{\l} = 0 \ {\rm for\
        large}\ \l
  \right\} \,.
\end{equation}
The precise meaning of $A_{\l} = 0$ for large $\l$ in (3.4) is: For
each $(A_{\l})_{\l > 0} \in {\bf A}_{x}(\cO)$ there is some positive
number $\l_A$ such that $A_{\l} = 0$ if $\l > \l_A$. (This requirement
is not central but turns out to be convenient.)
As in the case of testing-families of test-functions, we use the
notation $\lb A_{\l} \rb$ to denote the testing family $(A_{\l})_{\l >
  0}$. We note that ${\bf A}(\cO)$ is a $C^*$-algebra upon defining
the algebraic operations pointwise for each scaling-parameter $\l$
(i.e., $a \lb A_{\l} \rb + \lb B_{\l}\rb = \lb a A_{\l} + B_{\l} \rb$,
$\lb A_{\l} \rb \cdot \lb B_{\l} \rb = \lb A_{\l} \cdot B_{\l}\rb$, 
 $\lb A_{\l} \rb^* = \lb A_{\l}^* \rb$) and taking as $C^*$-norm
  $||\,\lb A_{\l}\rb\,|| = \sup_{\l > 0} \, ||\,A_{\l}\,||$. It is
 also clear that the map $\cO \to {\bf A}_x(\cO)$ is a net of
 $C^*$-algebras
 since the condition of isotony holds, i.e.\ ${\bf A}_x(\cO_1)
 \subset {\bf A}_x(\cO_2)$ for $\cO_1 \subset \cO_2$. Thus our
 testing-families will be the elements in the 
*-algebra ${\bf A}_x := \bigcup_{\cO}{\bf
    A}_x(\cO)$.
It should be noted that we do not take a closure of this set.
% On each ${\bf A}_x$ there acts a
% one-parametric automorphism group $\s_x^{\r}$, $\r > 0$, of scaling
% transformations (not continuous) defined by
% \begin{equation}
% \s_x^{\r}\lb A_{\l} \rb := \lb A_{\r\l} \rb \,.
% \end{equation}
% This implies that for all $x \in \bR^d$, $\r > 0$ and all double cone
% regions $\cO$ one has
% \begin{equation}
%  \s^{\r}_x({\bf A}_x(\cO)) = {\bf A}_x(\r\cO)\,.
% \end{equation}
Observe that 
\begin{equation}
 \a_y({\bf A}_x(\cO)) = {\bf A}_{x + y}(\cO)
\end{equation}
holds for all $x,y \in \bR^d$ and double cones $\cO$, where
\begin{equation}
 \a_y\lb A_{\l} \rb := \lb \a_y(A_{\l}) \rb \,.
\end{equation}

In the same way as we have used the testing-families in ${\bf F}_x$ to
probe the frequency behaviour of a distribution (a continuous linear
functional on the test-function space) infinitesimally close to the
point $x$ in coordinate space, we shall now employ the elements of the
scaling algebra ${\bf A}_x$ to analyse the frequency behaviour of a
continuous linear functional on $\cA$ close to $x$. To do so, we need
to introduce further notation. We will generically abbreviate an
element $(x_1,\ldots,x_n;k_1,\ldots,k_n) \in \bR^{dn} \times
(\bR^{dn}\backslash\{0\})$ by $(\xx;\kk)$. With this convention, $\kk
\cdot \yy$ denotes the scalar product $\sum_{j = 1}^n k_j\cdot y_j$
where in the sum appear the scalar products of the vectors $k_j$ and
$y_j$ in $\bR^d$; it should be borne in mind that the lower indices
here aren't coordinate indices. The integration measure 
$d^dy_1 \cdots d^dy_n$ will be abbreviated by $d\yy$. When
 $\lb A^{(1)}_{\l}\rb \otimes
\dots \otimes \lb A^{(n)}_{\l}\rb$
 is a simple tensor in ${\bf A}_{x_1}
\otimes \cdots \otimes {\bf A}_{x_n}$, then we denote this relation
simply by $\lb A^{(\xs)}_{\l}\rb \in {\bf A}_{\xs}$, understanding that
 $\xx = (x_1,\ldots,x_n)$.

\begin{Dfn}
Let $\varphi$ be a continuous linear functional on $\cA$, and let $n
\in \bN$. Then $ACS^n(\varphi)$, the {\em $n$-th order asymptotic
  correlation spectrum} of $\varphi$, is defined as the complement in
$\bR^{dn}\times(\bR^{dn}\backslash\{0\})$
 of all those $(\xx;\ks)$ which have the
following property: There is an $h \in \cD(\bR^{dn})$ with $h(0) = 1$
 and an open
neighbourhood $V$ of $\ks$ such that for each $\lb A^{(\xs)}_{\l} \rb \in
{\bf A}_{\xs}$ it holds that
\begin{equation}
\int \eilky\,h(\yy)\varphi(\a_{y_1}(A_{\l}^{(1)}) \cdots
\a_{y_n}(A_{\l}^{(n)}))\,d\yy \ = \ O^{\infty}(\l)\quad {\rm as}\quad \l
\to 0
\end{equation}
uniformly in $\kk \in V$.
\end{Dfn}
\noindent
{\it Remark.}  Since $\cA$ is an algebra, $\varphi$ can 
 be viewed as a linear map $\varphi_{\otimes}: \bigotimes^n \cA \to
\bC$,
$\varphi_{\otimes}(A_1 \otimes \cdots \otimes A_n) = \varphi(A_1
\cdots A_n)$. It is thus clear that the definition of $ACS^n(\varphi)$
for linear functionals $\varphi$ on $\cA$
generalizes, in an obvious manner,
to $ACS^n(\upsilon)$ for any  linear
functional $\upsilon : \bigotimes^n \cA \to \bC$ continuous on the
simple tensors.
\\[10pt]
We shall next collect a few immediate properties of $ACS^n(\varphi)$
which are reminiscent of corresponding properties of the wavefront
set.
\begin{Prop}
${}$ \\
(a) \quad (Analogue of Prop.\ 2.3.) If (3.7) holds for some choice of
$h$ and $V$, then it holds also with $\phi\cdot h$ in place of
$h$ for all $\phi \in C^{\infty}(\bR^{dn})$ and with $V$ replaced by
any open neighbourhood $V_1$ of $\ks$ fulfilling $\overline{V_1}
\subset V$.
 \\[6pt]
(b) \quad $ACS^n(\varphi)$ is a closed 
 subset of $\bR^{dn} \times (\bR^{dn} \backslash \{0\})$ which is
conic in the Fourier-space variables
 (this means that $(\xx;\ks) \in ACS^n(\varphi)$ iff
$(\xx;\mu \ks) \in ACS^n(\varphi)$ for all $\mu > 0$.)\\[6pt]
(c) \quad Translation covariance:  $(\xx;\ks) \in ACS^n(\varphi)$ if
and only if $(0;\ks) \in ACS^n(\varphi_{\xs})$ where for each $\xx \in
\bR^{dn}$, we define $\varphi_{\xs}: \bigotimes^n\cA \to \bC$ by
$\varphi_{\xs}(A_1 \otimes \cdots \otimes A_n) =
\varphi(\a_{x_1}(A_1) \cdots \a_{x_n}(A_n))$ (cf.\ the Remark
above).\\[6pt]
(d) \quad Suppose that the functional $\varphi$ is Hermitean, i.e.\
there holds $\varphi(A^*) = \overline{\varphi(A)}$, $A \in \cA$. Then
we have
\begin{equation}
 (\xx;\ks) \in ACS^n(\varphi) \Leftrightarrow (\bar{\xx};-\ks) \in
 ACS^n(\varphi) 
\end{equation}
 with $\bar{\xx} := (x_n,\ldots,x_1)$ for each $\xx =(x_1,\ldots,
 x_n)$.
\\[6pt]
(e) \quad $ACS^n(\varphi + \varphi') \subset ACS^n(\varphi) \cup
ACS^n(\varphi')$ holds for all continuous linear functionals
$\varphi,\varphi'$ on $\cA$.   
\end{Prop}
\begin{proof}
${}$\\
(a) Given $\lb A^{(\xs)}_{\l} \rb \in {\bf A}_{\xs}$, we
abbreviate:
\begin{equation}
\varphi_{\l}(\yy) := \varphi(\a_{y_1}(A_{\l}^{(1)}) \cdots
\a_{y_n}(A^{(n)}_{\l})) \,.
\end{equation}
Then we observe that that $\varphi_{\l}$, $\l > 0$, is a uniformly
bounded family of continuous functions, and thus $w_{\l}(\kk) :=
\widehat{h\varphi_{\l}}(\kk)$ is a family of smooth functions
satisfying the assumptions of Lemma 2.2($\b$). The statement follows
then by Lemma 2.2($\b$) upon noticing that $\phi$ may be replaced by
$\chi \phi$ for any $\chi \in \cD(\bR^{dn})$ with $\chi = 1$ on ${\rm
  supp}\,h$. \\[6pt]
(b) To show that $ACS^n(\varphi)$ is closed amounts to showing that, if
 $(\xx;\ks) \nin ACS^n(\varphi)$, then  there are
open neighbourhoods $U$ of $\xx$ and $W$ of $\ks$ so that $(\xx';\ks')
\nin ACS^n(\varphi)$ for all $\xx' \in U$ and $\ks' \in W$.

 So let $(\xx;\ks) \nin ACS^n(\varphi)$. This means that (3.7) holds
 with suitable choices of $V$  and $h$, where the function $h$ is
 greater than some strictly positive constant in some neighbourhood
 $N$ of $\yy =0$.
  Let $N_1$ and $N_2$ be two other
neighbourhoods of $0 \in \bR^{dn}$ such that $N_1 + N_2\subset
N$. Then choose some $h_1 \in \cD(\bR^{dn})$ with ${\rm supp}\,h_1
\subset N_1$ and such that $h_1(0) = 1$. Now for all $\lb A_{\l}^{\xs}
\rb \in {\bf A}_{\xs}$ and all $\yy' \in N_2$, one obtains
\begin{eqnarray} \lefteqn{
 \int \eilky h_1(\yy) \varphi(\a_{y_1 - y_1'}(A_{\l}^{(1)}) \cdots 
\a_{y_n - y_n'}(A_{\l}^{(n)}))\,d\yy}  \\
 & = & 
 {\rm e}^{- i \l^{-1} \mbox{\scriptsize\boldmath $k \cdot y'$}}\!\!
 \int \eilky h_1(\yy + \yy') \varphi(\a_{y_1}(A_{\l}^{(1)}) \cdots 
 \a_{\l}(A_{\l}^{(n)}) )\,d\yy = O^{\infty}(\l) \ \ {\rm as} \ \  \l \to
 0 \nonumber
\end{eqnarray}
uniformly in $\kk \in V_1$ for some open neighbourhood $V_1$ of $\ks$,
 since by construction, $h_1$ has the
property that there is for each $\yy' \in N_1$ a
$\phi_{\mbox{\scriptsize\boldmath $y'$}} \in C^{\infty}(\bR^{dn})$
with $(\phi_{\mbox{\scriptsize\boldmath $y'$}}h)(\,.\,) = h_1(\,.\,)$.
By the covariance property (3.5), this implies that we have for all
$\xx'$ in the open neighbourhood $U =\{\xx -\yy':\yy' \in N_2\}$ of
$\xx$ 
\begin{equation}
\int \eilky h_1(\yy)\varphi(\a_{y_1}(A_{\l}^{(1)}) \cdots 
 \a_{y_n}(A_{\l}^{(n)}) ) = O^{\infty}(\l) \quad {\rm as} \quad \l \to 0
\end{equation}
uniformly in $\kk \in V_1$ whenever $\lb A_{\l}^{(\xs')} \rb \in {\bf
  A}_{\xs'}$. Since $V_1$ is an open neighbourhood of $\ks$, it is
clear that we can also find an open neighbourhood $W$ of $\ks$ such
that each $\ks' \in W$ possesses some open neighbourhood
$V_{\mbox{\scriptsize\boldmath $\xi'$}} \subset V_1$. Hence, for each
$\xx' \in U$ and $\ks' \in W$ condition (3.11) holds for all 
$\lb A_{\l}^{(\xs')} \rb \in {\bf
  A}_{\xs'}$ uniformly in $k \in V_{\mbox{\scriptsize\boldmath
    $\xi'$}}$. This shows $(\xx';\ks') \nin ACS^n(\varphi)$ for $\xx'
\in U$, $\ks' \in W$.

Next we show that $(\xx;\ks) \nin ACS^n(\varphi)$ implies 
$(\xx;\mu\ks) \nin ACS^n(\varphi)$, thus establishing the conicity of
$ACS^n(\varphi)$ in the Fourier-space variables. If $(\xx;\ks) \nin
ACS^n(\varphi)$, then we have for all 
 $\lb A_{\l}^{(\xs)} \rb \in {\bf A}_{\xs}$, using the notation
(3.9), 
\begin{equation}
 \int \eilky h(\yy)\varphi_{\l}(\yy) \,d\yy = O^{\infty}(\l) \quad
 {\rm as} \quad \l \to 0
\end{equation}
uniformly in $\kk \in V$ for suitable $h$. Setting $\r = \mu^{-1} >
0$, we replace on the left hand side of the last equation the
parameter $\l$ by $\r\l$. Denoting $\lb A_{\r\l}^{(\xs)}\rb$ by
$\lb A_{\l}'{}^{(\xs)}\rb$, this yields  
\begin{equation}
\int {\rm e}^{-i \l^{-1}\mu \mbox{\scriptsize\boldmath $k \cdot y$}}
 h(\yy) \varphi(\a_{y_1}(A'{}_{\l}^{(1)}) \cdots
 \a_{y_n}(A'{}_{\l}^{(n)}))\, d\yy = O^{\infty}(\l)\quad {\rm as} \quad \l
 \to 0
\end{equation}
uniformly in $\kk \in V$ for all $\lb A'{}_{\l}^{(\xs)} \rb \in {\bf
  A}_{\xs}$, since ${\bf A}_x$ is invariant under the
scale-transformations $\lb A_{\l} \rb \mapsto \lb A_{\r\l}\rb$, $\r >
0$. Hence $(\xx;\mu \ks) \nin
ACS^n(\varphi)$.
\\[6pt]
(c) This is simply a consequence of (3.5).
\\[6pt]
(d) The claimed property is easily verified by inspection. (e) is obvious. 
\end{proof}
%%%%%%%%%%%%%%%%%%%%%%%%%%%%%%%%%%%%%%%%%%%%%%%%%%
\section{The ACS in algebraic quantum field theory}
%%%%%%%%%%%%%%%%%%%%%%%%%%%%%%%%%%%%%%%%%%%%%%%%%%
\setcounter{equation}{0}
%%%%%%%%%%%%%%%%%%%%%%%%%%%%%%%%%%%%%%%%%%%%%%%%%%
In Section 3, we have described a ``theory'' just by an
inclusion-preserving map $\cO \to \cA(\cO)$ assigning $C^*$-algebras
to double cone regions  together with with a covariant, strongly
continuous action $\{\a_x\}_{x \in \bR^d}$ of the translation group
by automorphisms. Now we shall add more structure which is
characteristic of quantum field theory proper --- like locality and
the spectrum condition --- and investigate what properties of
$ACS^n(\varphi)$ for functionals or states $\varphi$ on $\cA =
\overline{\bigcup_{\cO} \cA(\cO)}^{C^*}$ ensue. We will also deduce
some consequences resulting from imposing certain ``upper bounds'' on
the shape of $ACS^n(\varphi)$.

The first relevant assumption we add to a theory
$(\cO \to \cA(\cO),\{\a_x\}_{x \in \bR^d})$ with the properties
listed at the beginning of Section 3.1 is:
\\[6pt]
$(SC)$  The theory 
$(\cO \to \cA(\cO),\{\a_x\}_{x \in \bR^d})$ is given in a covariant
representation satisfying the {\it spectrum condition}. That means, $\cA$ is
an algebra of bounded operators on a Hilbertspace $\cH$, and there is
a weakly continuous representation $\bR^d \owns x \mapsto U(x)$ of the
translation group by unitary operators on $\cH$ such that $\a_x(A) =
U(x)AU(x)^{-1}$, $x \in \bR^d$, $A \in \cA$. Moreover, it holds that
the spectrum of $P = (P_{\mu})_{\mu = 0,\ldots,d-1}$, the generator of
of $U(x) = {\rm e}^{iP_{\mu}x^{\mu}}$, $x = (x^{\mu})_{\mu =
  0,\ldots,d-1}$, is contained in the $d$-dimensional closed forward
lightcone $\overline{V}_+$.
(The existence of a vacuum vector is not assumed here.)
Note that in the presence of $(SC)$ the unitary group
$U(x)$, $x \in \bR^d$, may be chosen to be contained in $\cA''$
\cite{Bor1,Arv} (see also \cite[Chp. II]{Bor2} and
\cite[Prop. 2.4.4]{Sak})
 and we shall henceforth assume that such a choice has been made.
\\[6pt]
The next point is to define a class of states, or functionals, on
$\cA$ which we wish to investigate. In the presence of $(SC)$, these
are the continuous functionals on $\cA$ which are normal, i.e.\ they
admit a normal extension to $\cA''$. Furthermore, we demand that the
functionals are ``$C^{\infty}$ for the energy''. One can define
several versions of this property. To present ours, we use the
standard notation $D_x^{\b} = i\frac{\partial^{\b_0}}{\partial x^{0}}
\cdots i\frac{\partial^{\b_{d-1}}}{\partial x^{d-1}}$ for iterated
partial derivatives, where $\b =(\b_0,\ldots,\b_{d-1}) \in \bN^d_0$ is a
multi-index. 
\\[6pt]
$(s - C^{\infty})$  A continuous, normal functional $\varphi$
on $\cA$ is called {\it strongly} $C^{\infty}$ $(s - C^{\infty})$ if the
partial derivatives
\begin{equation} 
                 D_x^{\b}D_y^{\g}\varphi(U(x)AU(-y))\,, \quad A \in
                 \cA,\ x,y \in \bR^d\,,
\end{equation}
exist for all multi-indices $\b,\g$  and induce normal functionals on
$\cA$. 
\\[6pt]
{\it Remarks.} (i) Standard examples of strongly $C^{\infty}$
functionals may be obtained from $C^{\infty}$ vectors for the energy,
i.e.\ such vectors $\psi \in \cH$ which are contained in the domain of
$(P_0)^N$ for all $N \in \bN$. In view of the spectrum condition,
$\psi$ lies then also in the domain of any power of $P_{\mu}$, $\mu =
1,\ldots,d-1$. Any two $C^{\infty}$ vectors $\psi',¸\psi$ give rise to
a normal, strongly $C^{\infty}$ functional $\varphi(A) = \langle
\psi',A\psi\rangle$, $A \in \cA$.
\\[6pt]
(ii) An equivalent way of expressing the $s - C^{\infty}$ property ---
which we will make use of --- is the following, as may easily be
checked: When we denote by $F_{\varphi,A}$ the Fourier transform of
the function $(x_1,x_2) \mapsto \varphi(U(x_1)AU(-x_2))$ (this
function is a tempered distribution, and so is its Fourier transform),
it follows that for any $N \in \bN$ and any $\phi \in \cS(\bR^d \times
\bR^d)$ there is some constant $c > 0$ so that 
\begin{equation}
 |\,\phi * F_{\varphi,A}(k_1,k_2)\,|(1 + |k_1| + |k_2|)^N \le
 c\cdot||\,A\,||\,, \quad A \in \cA\,, \ k_1,k_2 \in \bR^d\,.
\end{equation}
\par\smallskip\noindent
The formulation of the subsequent result is preceded by a list of
notational conventions: For a continuous linear functional $\varphi$
on $\cA$ and we write
\footnote{The closure in (4.4) is understood in the set
  $\bR^d\backslash \{0\}$.}
\begin{eqnarray}
 ACS_{\xs}^n(\varphi) & := & \{\kk \in \bR^{dn}\backslash
 \{0\}:(\xx;\kk) \in ACS^n(\varphi)\}\,, \quad \xx \in \bR^{dn}\,, \\
\p_2ACS^n(\varphi) & := & [\bigcup_{\xs \in \bR^{dn}}
ACS^n_{\xs}(\varphi)]^-\,. 
\end{eqnarray}
We denote $n$-tupels of vectors in $\bR^d$ again by $\kk =
(k_1,\ldots,k_n)$, and set
\begin{equation}
 k^{[j]} := \sum_{i = j}^n k_i \,, \quad j = 1,\ldots,n\,.
\end{equation}
Then we define the set
\begin{equation}
 \cV_n := \left\{ (k_1,\ldots,k_n) \in \bR^{dn}: k^{[j]} \in
   \overline{V}_+,\ j = 2,\ldots,n,\ k^{[1]} = 0\right\} \,.
\end{equation}
Thus $\cV_n$ coincides with the bound for the support of the
Fourier-transformed $n$-point vacuum expectation values in quantum
field theory \cite{SW}.
\begin{Prop}
Suppose that the theory satisfies the spectrum condition $(SC)$, and
let $\varphi$ be a continuous, normal functional on $\cA$. If
$\varphi$ is strongly $C^{\infty}$ then
$$ \p_2ACS^n(\varphi) \subset \cV_n\backslash\{0\}\ \ 
 \quad {\rm for\ all}\ n \in
\bN\,.$$
\end{Prop} 
\begin{proof}
$\cV_n$ is a closed set in $\bR^{dn}$. We set $\cX_n :=
\bR^{dn}\backslash \cV_n$ which is open both in $\bR^{dn}$ and
$\bR^{dn} \backslash \{0\}$. To prove the Proposition, it suffices to
show that given $\xx \in \bR^{dn}$ and $\ks \in \cX_n$, there is an
open neighbourhood $V$ of $\ks$ and some $h \in \cD(\bR^{dn})$ with
$h(0) = 1$ so that for all $\lb A_{\l}^{(\xs)} \rb \in {\bf A}_{\xs}$
there holds
\begin{equation}
\int \eilky h(\yy) \varphi_{\l}(\yy)\,d\yy = O^{\infty}(\l) \quad {\rm
  as} \quad \l \to \infty
\end{equation}
uniformly in $\kk \in V$; our by now familiar abbreviation
\begin{equation}
\varphi_{\l}(\yy) = \varphi(\a_{y_1}(A_{\l}^{(1)}) \cdots
\a_{y_n}(A_{\l}^{(n)})) 
\end{equation}
will be recalled.

In the following, there will often appear $n + 1$-tupels of vectors in
$\bR^{d}$ which will be denoted
\begin{equation}
 \zul = (\zz,z_{n+1}) = (z_1,\ldots,z_{n+1})\,.
\end{equation}
To exploit the spectrum condition, it is customary to pass from the
variable $\yy$ in (4.8) to relative variables $z_1 = y_1$, $z_2 = y_2
- y_1$, $z_n = y_n - y_{n-1}$.
In this way one obtains, upon setting
\begin{equation}
\Psi_{\l}(\zul) := \varphi(U(z_1)A^{(1)}_{\l}U(z_2)A_{\l}^{(2)}U(z_3)
\cdots U(z_{n})A^{(n)}_{\l}U(-z_{n+1}))\,,
\end{equation}
that
\begin{equation}
\Phi_{\l}(\zz) := \Psi_{\l}(\zz,-\sum_{j = 1}^n z_j) =
\varphi_{\l}(\yy)\,,
\end{equation}
and similarly, with $g(\zz) = h(z_1,z_1 + z_2,\ldots,\sum_{j=1}^n
z_j)$,
\begin{equation}
\int {\rm e}^{- i\l^{-1}\sum_{j=1}^n k^{[j]}\cdot
  z_j}g(\zz)\Phi_{\l}(\zz)\,d\zz = \int \eilky
h(\yy)\varphi_{\l}(\yy)\,d\yy\,.
\end{equation}
Now let $\G_n := \{0\} \times \overline{V}_+ \times \cdots \times
\overline{V}_+ \subset (\bR^d)^n$ (the set $\overline{V}_+$ appears
$n-1$ times) and $R_n := (\bR^d)^n \backslash \G_n$. Observe that we
have
\begin{equation}
(k^{[1]},\ldots,k^{[n]}) \in \G_n \Leftrightarrow (k_1,\ldots,k_n) \in
\cV_n\,.
\end{equation}
We will now demonstrate that given any conic subset $E \subset R_n$
which is closed in $\bR^{dn}\backslash \{0\}$, any $g \in
\cD(\bR^{dn})$ and any $\lb A_{\l}^{(\xs)}\rb \in {\bf A}_{\xs}$ for
arbitrary $\xx \in \bR^{dn}$, one can find for each $N \in \bN$ some
number $c>0$ such that 
\begin{equation}
\sup_{\l}\,\sup_{\kks \in E} |\widehat{g\Phi_{\l}}(\kk)| (1 + |\kk|)^N
\le c\,.
\end{equation}
This property can be seen to imply, in view of (4.5,12,13), the
required relation (4.7).
To prove (4.14) one first observes that the assumptions entail the
following properties of $\widehat{\Psi_{\l}}$, the Fourier-transform
of $\Psi_{\l}$: Roughly speaking,
$\widehat{\Psi_{\l}}(k_1,\ldots,k_{n+1})$ is rapidly decreasing in the
first and last entries $k_1$ and $k_{n+1}$ (implied by the $s -
C^{\infty}$ property), and has support in $\overline{V}_+$ with
respect to each of the remaining variables $k_2,\ldots,k_n$ (implied
by $(SC)$). Moreover, these properties are uniform in $\l$. But we
must take into account that $\widehat{\Psi_{\l}}$ is actually a
distribution, requiring a slightly different formulation of these
properties.
So let $\G'_{n+1} := \{0\} \times \overline{V}_+ \times \cdots \times
\overline{V}_+ \times \{0\} = \G_n \times \{0\} \subset
(\bR^d)^{n+1}$ (the set $\overline{V}_+$ appears again $n -1$ times)
and $R_{n+1}' := (\bR^d)^{n+1}\backslash \G_{n+1}'$. What we will
show is that given any conic subset $E' \subset R'_{n+1}$ which is
closed in $\bR^{d(n+1)}\backslash\{0\}$, any $\chi \in
\cD(\bR^{d(n+1)})$ and any $\lb A_{\l}^{(\xs)}\rb \in {\bf A}_{\xs}$
one can find for every $N \in \bN$ some constant $c' > 0$ with
\begin{equation}
\sup_{\l}\,\sup_{\kul \in E'} |\widehat{\chi\Psi_{\l}}(\kul)|(1 +
|\kul|)^N \le c'\,.
\end{equation}
Let us point out how this property entails (4.14). Define $Q : \bR^{dn}
\to \bR^{d(n+1)}$ by $Q(\zz) := (\zz,-\sum_{j =1}^nz_j)$. The
derivative $DQ$ of this map is constant, and its transpose is given by
${}^t(DQ)\kul = (k_1 - k_{n+1},\ldots,k_n - k_{n+1})$. The set
$$ N_Q = \{ (Q(\zz),\kul)\in \bR^{d(n+1)} \times \bR^{d(n +1)}:
{}^t(DQ)\kul = 0\} $$
is therefore  contained in $\bR^{d(n+1)} \times \D_{n+1}$ where
$\D_{n+1}$ is the total diagonal in $(\bR^d)^{n+1}$. Since $\D_{n+1}
\cap (\G'_{n+1}\backslash \{0\}) = \emptyset$, we see that
$$ N_Q \cap \left[ \bR^{d(n+1)} \times (\G'_{n+1}\backslash
  \{0\})\right] = \emptyset\,. $$
Observe also that ${}^t(DQ)(\G_{n+1}'\backslash \{0\}) =
 \G_n\backslash \{0\}$. Thus we can apply
Theorem 8.2.4 in \cite{Hor1} which says, for our situation, that (4.15)
implies for any conic subset $E$ of $\bR^{dn} \backslash \{0\}$ with
$\overline{E} \subset R_n$ the relation
\begin{equation}
 \sup_{\l}\,\sup_{\kks \in E} |((\widehat{\chi\Psi_{\l})\lcrc
 Q})(\kk)| (1 + |\kk|)^N < c_N
\end{equation}
for all $N \in \bN$ with suitable constants $c_N > 0$. Since
$\Psi_{\l} \lcrc Q = \Phi_{\l}$, one deduces (4.14) from (4.16).

So we are left with having to prove relation (4.15). The proof
proceeds by a variation of more or less standard arguments which can
be found in slightly different forms in the literature, e.g.\
\cite[Chp. VIII]{Hor1}. To begin with, we have that
$\sup_{\l}\,||\,\Psi_{\l}\,||_{\infty} < a$ for some $a >0$, thus
$\widehat{\psi}\Psi_{\l} \in L^1$ and $\sup_{\l}\,||\,\psi *
\widehat{\Psi_{\l}}\,||_{\infty} \le b$ for some $b > 0$ whenever
$\psi \in \cS(\bR^{d(n+1)})$. Consequently, if $\r_{\m}(\kul) =
\r(\kul/\m)$, $\kul \in \bR^{d(n+1)}$, $\mu > 0$, where $\r \in
\cD(\bR^{d(n+1)})$, $0 \le \r \le 1$ and $\r$
 is equal to 1 on an arbitrary open ball 
containing the origin, we obtain for any $\phi \in \cS(\bR^{d(n+1)})$,
 any $N \in \bN$ and any $\eta > 1$
\begin{equation}
\sup_{\l}\,\sup_{\kul \in \bR^{d(n+1)}} |\psi *
\widehat{\Psi_{\l}}(\t_{\eta\kul}(\phi - \r_{|\kul|}\phi))| (1 + |\kul|)^N
\le C_N
\end{equation}
for suitable $C_N > 0$. To see this, let $s > 0$ be the
radius of the  open ball around the origin where
$\r =1$. Then consider
\begin{eqnarray*}
\lefteqn{
|\psi *
\widehat{\Psi_{\l}}(\t_{\eta\kul}(\phi - \r_{|\kul|}\phi))| (1 +
|\kul|)^N}\\
& \le & b \int |\phi(\kul')-(\r_{|\kul|}\phi)(\kul')|(1 + |\kul|)^N
\,d\kul'\\
& \le & b' \int_{|\kul'| \ge s|\kul|} |\phi(\kul')|(1 +
|\kul|)^N\,d\kul' \\
& \le & b'' \int_{|\kul'| \ge s|\kul|} \frac{1}{(1 +
  |\kul'|)^{M + N}}(1 + |\kul|)^N \,d\kul' \ \le \ C_N \,;
\end{eqnarray*}
obviously this chain of estimates holds upon suitable choice of positive
constants $b',b'',M$ and $C_N$.

Now let $\ksul = (\xi_1,\ldots,\xi_{n+1}) \in R_{n+1}'$. We
distinguish two cases:
\begin{itemize}
\item[(i)] $|\xi_1| + |\xi_{n+1}| > 0$
\item[(ii)] $\xi_1 = \xi_{n+1} = 0$
\end{itemize}
{\it Case} (i). One infers that there is some open conic
neighbourhood $E'_{\ksul}$ of $\ksul$ with the property 
\begin{equation}
 \vartheta(|k_1| + |k_{n+1}|) \ge |\kul|\,, \quad \kul \in
 E'_{\ksul}\,,
\end{equation}
for some suitable $\vartheta > 0$. Let $\chi = \chi_1 \otimes \cdots
\otimes \chi_{n+1}$ with $\chi_j \in \cS(\bR^d)$, $j = 1,\ldots,n+1.$
Recalling the notation introduced in Remark (ii) above we find
\begin{equation}
\widehat{\chi}*\widehat{\Psi_{\la}}(\kul) = (\widehat{\chi_1} \otimes
\widehat{\chi_{n+1}} ) * F_{\varphi,B_{\l,\kul}}(k_1,k_{n+1})\,,
\end{equation}
where
\begin{equation}
B_{\l,\kul} := \int {\rm e}^{-i(k_2z_2 + \cdots + k_nz_n)}\chi_2(z_2)
\cdots \chi_n(z_n)\cdot A_{\l}^{(1)}U(z_2) \cdots
U(z_n)A^{(n+1)}_{\l}\,d^dz_2 \cdots d^dz_n\,.
\end{equation}
Actually $B_{\l,\kul}$ is independent of $k_1$ and $k_{n+1}$ and
$||\,B_{\l,\kul}\,|| < {\rm const.}$, thus we may deduce from (4.18)
together with (4.2) that
\begin{eqnarray}
\lefteqn{\sup_{\l}\,\sup_{\kul \in E'_{\ksul}}\,|\widehat{\chi}*
  \widehat{\Psi_{\l}}(\kul)|(1 + |\kul|)^N } \\
& \le & \sup_{\l}\,\sup_{\kul \in E'_{\ksul}}\,|(\widehat{\chi_1}
\otimes \widehat{\chi_{n+1}})* F_{\varphi,B_{\l,\kul}}(k_1,k_{n+1})|\,
\left[1 + \vartheta(|k_1| + |k_2|) \right]^N \le C_N \nonumber
\end{eqnarray}
for all $N \in \bN$ with some $C_N > 0$.
\\[6pt]
{\it Case} (ii). In this case there is a conic open
neighbourhood $E''_{\ksul}$ of $\ksul$ with $(k_2,\ldots,k_n) \in
(\bR)^{n-1} \backslash \overline{V}_+ \times \cdots \times
\overline{V}_+$ for all $\kul \in E''_{\ksul}$. We may suppose that
$E_{\ksul}'' = \bR^+(\eta'\ksul + \overline{\cO_1})$ where $\cO_1$ is
the unit ball around the origin in $\bR^{d(n+1)}$ and $\eta'$ is some
suitable number greater than 1. Now let $\r \in \cD(\bR^{d(n+1)})$, $0
\le \r \le 1$, and such that $\r$ has support in $\cO_1$ and is equal
to 1 on $\frac{1}{2}\cO_1$. Moreover, let $E'_{\ksul} =
\bR^+(2\eta'\ksul + \overline{\cO_1})$, and let $\eta > 4\eta'$. Then
it follows that for all $\kul \in E'_{\ksul}$ and all $\phi \in
\cS(\bR^{d(n+1)})$ and  $\psi \in \cD(\bR^{d(n+1)})$ one has
\begin{equation}
 {\rm supp}(\t_{\eta \kul}(\r_{|\kul|}\phi * {}^r\!\psi)) \subset
 E''_{\ksul}
\end{equation}
as soon as $|\kul|$ is large enough (depending on the support of
$\psi$).

Assuming now that $\phi = \phi_1 \otimes \cdots \otimes \phi_{n+1}$
and $\psi = \psi_1 \otimes \cdots\otimes \psi_{n+1}$ with $\phi_j \in
\cS(\bR^d)$ and $\psi_j \in \cD(\bR^d)$, the spectrum condition $(SC)$
implies that, if $\kul$ is contained in $E_{\ksul}'$ and $|\kul|$
sufficiently large, then
\begin{equation}
\widehat{\Psi_{\l}}(\t_{\eta \kul}(\r_{|\kul|}\phi* {}^r\!\psi)) = 0
\end{equation}
holds for all $\l > 0$ because of (4.22) and since each $\kul \in
E''_{\ksul}$ has $(k_2,\ldots,k_n) \subset (\bR^d)^{n-1} \backslash
\overline{V}_+ \times \cdots \times \overline{V}_+$. In view of
(4.17) we therefore obtain that 
\begin{eqnarray} \lefteqn{
\sup_{\l}\,\sup_{\kul \in E'_{\ksul}}\, |{}^r\!\phi*\psi *
\widehat{\Psi_{\l}}(\eta\kul)|\,(1 + |\kul|)^N} \\
& \le & \sup_{\l}\,\sup_{\kul \in E'_{\ksul}}\,|\psi *
\widehat{\Psi_{\l}}(\t_{\eta \kul}(\phi - \r_{|\kul|}\phi))|\,(1 +
|\kul|)^N
\nonumber \\
&  & +\ \sup_{\l}\,\sup_{\kul \in E'_{\ksul}}\,|
\widehat{\Psi_{\l}}(\t_{\eta \kul}(\r_{|\kul|}\phi* {}^r\!\psi)|\,(1 +
|\kul|)^N
\nonumber \\
& \le & C_N \nonumber
\end{eqnarray}
holds for each $N \in \bN$ with some suitable $C_N > 0$.
\\[6pt]
Now every open conic subset $E' \subset R'_{n+1}$ with $\overline{E'}
\subset R'_{n+1}$ can be covered by finitely many conic neighbourhoods
of the type $E'_{\ksul}$, $\ksul \in R'_{n+1}$, corresponding to the
cases (i) or (ii) just considered. Relation (4.15) is thus proved by
(4.21) and (4.24) apart from a remaining step which is to pass from the special
functions $\chi = \chi_1 \otimes \cdots \otimes \chi_{n+1}$ and $\chi
= \widehat{\phi_1}\widehat{{}^r\!\psi_1} \otimes \cdots \otimes
\widehat{\phi_{n+1}}\widehat{{}^r\!\psi_{n+1}}$, which we considered in the
cases (i) and (ii), respectively, to generic $\chi \in
\cD(\bR^{d(n+1)})$. The argument showing this is, however, standard
\cite[Lemma 8.1.1]{Hor1}; it is in essence contained in the proof of
Prop.\ 2.3,  and we therefore skip the details.
\end{proof}

The next result which we list is a simple observation combining the
assumption that the Fourier-space component of the $ACS$ is confined
within a salient cone with the condition of locality, i.e.\ the
property that elements of $\cA(\cO_1)$ and $\cA(\cO_2)$ commute once
the localization regions are acausally separated. Then it follows that
certain elements of $\bR^{dn} \times (\bR^{dn} \backslash \{0\})$ are
absent from $ACS^n(\varphi)$ for hermitean functionals $\varphi$
on $\cA$. Such statements are known for the wavefront sets of Wightman
distributions (they appear e.g.\ implicitly in
\cite{BFK}). Nevertheless it seems appropriate to put the simple
argument on record here.

We begin by fixing the condition of locality which is motivated by
Einstein causality (signals propagate with at most the velocity of
light), cf.\ \cite{HK,Haag}.
\\[6pt]
$(L)$ The theory $(\cO \to \cA(\cO), \{\a_x\}_{x \in \bR^d})$ is said
to fulfill {\it locality} if\footnote{here $[\cA(\cO_1),\cA(\cO_2)]=
  \{A_1A_2 - A_2A_1 : A_j \in \cA(\cO_j),\ j = 1,2\}$}
\begin{equation}
 [\cA(\cO_1),\cA(\cO_2)]= \{0\}
\end{equation}
whenever the regions $\cO_1$ and $\cO_2$ are acausally related, i.e.\
there is no causal curve joining $\cO_1$ and $\cO_2$ (equivalently,
$\cO_1 \cap \left[\pm \overline{V}_+ + \cO_2\right] = \emptyset$).
\\[6pt]
{\it Remark.} This form of locality is sometimes referred to as {\it
  spacelike commutativity}. There are theories (e.g.\ conformally
covariant theories) which additionally fulfill {\it timelike
  commutativity} which means that (4.25) holds provided there is no
timelike curve joining $\cO_1$ and $\cO_2$. Thus, in a theory
fulfilling both spacelike and timelike commutativity one has (4.25) as
soon as there is no lightlike line connecting $\cO_1$ and $\cO_2$.
\\[6pt]
We shall say that an $n$-tupel $\xx = (x_1,\ldots,x_n) \in \bR^{dn}$
is {\it properly acausal} ({\it properly non-lightlike}) if there is no
causal (lightlike) curve joining any pair of points $x_j$ and $x_i$
for $i \ne j$, $i,j = 1,\ldots,n$. A (maximal) {\it salient cone} $\cW$ in
$\bR^{dn} \backslash \{0\}$ is, by definition, a conic subset of
$\bR^{dn} \backslash \{0\}$ such that $\cW \cap -\cW = \emptyset$.

With this notation, we arrive at:

\begin{Prop}
Suppose that the theory $(\cO \to \cA(\cO), \{\a_x\}_{x \in \bR^d})$
fulfills locality $(L)$ and let $\cW$ be a closed salient cone in
$\bR^{dn} \backslash \{0\}$. Then for any continuous hermitean
functional $\varphi$ on $\cA$ the conditions
$$ ACS_{\xs}^n(\varphi) \subset \cW \quad {\rm and} \quad \xx \ \,
{\rm properly\ \, acausal} $$
imply \ \  $ACS^n_{\xs}(\varphi) = \emptyset$.

The analogous statement  with ``properly acausal'' replaced by
``properly non-lightlike'' holds for a theory fulfilling both spacelike
and timelike commutativity.
\end{Prop}
\begin{Cor}
For a theory satisfying locality $(L)$ and spectrum condition $(SC)$,
it holds that 
$$ ACS_{\xs}^n(\varphi) = \emptyset $$
if $\varphi$ is a strongly $C^{\infty}$, continuous hermitean
functional on $\cA$ and $\xx$ is properly acausal. (Again there holds
the sharpened version of this statement with ``$\xx$ properly
non-lightlike'' for a theory satisfying also timelike commutativity.)
\end{Cor}
\begin{proof}
The Corollary follows simply from Propositions 4.1 and 4.2 since
it is elementary to check that the set $\cV_n \backslash \{0\}$ is a
closed salient cone in $\bR^{dn}\backslash \{0\}$. To prove Prop.\
4.2, let $\lb A_{\l}^{(\xs)}\rb \in {\bf A}_{\xs}$ for $\xx$ properly
acausal. As a consequence of the assumptions, each $\ks \in
\bR^{dn}\backslash\cW$ possesses an open neighbourhood $V$ so that
\begin{equation}
 \int \eilky 
h(\yy)\varphi(\a_{y_1}(A_{\l}^{(1)})\cdots
 \a_{y_n}(A_{\l}^{(n)})) \,d\yy = O^{\infty}(\l) \quad {\rm as} \quad
 \l \to 0 
\end{equation}
holds uniformly for $\kk \in V$ with some suitable $h \in
\cD(\bR^{dn})$, $h(0) = 1$. In view of Prop.\ 3.2(a) we may assume that
$h$ is real and that
the diameter of the support of $h$ is smaller than $\frac{1}{3}\min_{i
  \ne j}\,|x_i - x_j|$. Using Proposition 3.2(d) it follows that $h$
and $V$ may be chosen in such a way that one also has
\begin{equation}
 \int {\rm e}^{-i \l^{-1}(-\mbox{\scriptsize\boldmath $k$}) \cdot
 \mbox{\scriptsize\boldmath $y$}} 
h(\yy)\varphi(\a_{y_n}(A_{\l}^{(n)})\cdots
 \a_{y_1}(A_{\l}^{(1)})) \,d\yy = O^{\infty}(\l) \quad {\rm as} \quad
 \l \to 0 
\end{equation}
uniformly in $\kk \in V$, but since the $\a_{y_j}(A_{\l}^{(j)})$, $\yy
\in {\rm supp}\,h$, $j = 1,\ldots,n$, pairwise commute for
sufficiently small $\l$ we conclude that the left hand side of (4.26)
equals the left hand side of (4.27) with $\kk$ replaced by $-\kk$. This
amounts to saying that under the stated assumptions, $ACS^n_{\xs}(\varphi)
\subset \cW$ entails $ACS^n_{\xs}(\varphi) \subset - \cW$ and thus
$ASC_{\xs}^n(\varphi) = \emptyset$ since $\cW$ is a salient cone in
$\bR^{dn} \backslash \{0\}$.
\end{proof}
In a further step we study the relation of properties of the $ACS$ to
properties of the scaling limit of the given theory in the sense of
\cite{BV1}. To do so we have to begin with some preparation, i.e.\ we
need to summarize some parts of the notions developed in
\cite{BV1}.  Let  
$(\cO \to \cA(\cO), \{\a_x\}_{x \in \bR^d})$ be a theory as in
Sec.\ 3, so that $\cO \to \cA(\cO)$ is just a net of $C^*$-algebras
over $d$-dimensional Minkowski-spacetime on which the translations act
as a $C^*$-dynamical system. We suppose such a theory is now given and
keep it fixed; it will be referred to as {\it the given theory} or
also {\it the underlying theory}. Note that more is not assumed
presently about the given theory (like e.g.\ ($SC$), ($L$) or the
existence of a vacuum state).
 
In \cite{BV1}, the scaling algebra associated with a given theory was
introduced as a means for the analysis of the theory's short distance
behaviour. The local scaling algebras at a point $x \in \bR^d$,
denoted by $\cAu_x(\cO)$, are defined as the $C^*$-subalgebras of ${\bf
  A}_x(\cO)$ formed by all the testing families $\lb A_{\l}\rb$ with
the property
\begin{equation}
\sup_{\l > 0}\, ||\,\a_{\l x}(A_{\l}) - A_{\l}\,|| \to 0 \quad {\rm
  for}\quad \l \to 0\,.
\end{equation}
This property contrains the growth of the energy-momentum transferred
by $A_{\l}$ as $\l \to 0$; see \cite{BV1} for discussion. The scaling
algebra at $x$ is then defined as $\cAu_x :=
\overline{\bigcup_{\cO}\cAu_x(\cO)}^{C^*}$.  
It is now useful to adopt the notation (cf.\ \cite{BV1}) to write
$\Au$ instead of $\lb A_{\l} \rb$ for testing families in
$\cAu_x$. In other words, the function $\Au : \bR^+ \to \cA$ denotes
the testing family $(A_{\l})_{\l > 0}$ in $\cAu_x$, and $\Aul$ stands
for the value of that function evaluated at some argument $\l \in
\bR^+$. Then we define as in \cite{BV1} the action of the
translations lifted to $\cAu_x$ by
\begin{equation}
(\ayu(\Au))_{\l} := \a_{\l y}(\Aul)\,, \quad y \in\bR^d\,,\ \l > 0\,,
\ \Au \in \cAu_x\,.
\end{equation}
One easily checks that $\ayu$ is a $C^*$-automorphism of the
$C^*$-algebra $\cAu_x$ which acts as translation on the local scaling
algebras at $x$, that is,
\begin{equation}
 \ayu(\cAu_x(\cO)) = \cAu_x(\cO + y)
\end{equation}
holds for all double cone regions $\cO$ and all $x,y \in
\bR^d$. Moreover, as a consequence of the condition (4.28)
 it follows that
$\{\ayu\}_{y \in \bR^d}$ is a $C^*$-dynamics on each $\cAu_x$, $x \in
\bR^d$.

Fixing some $x \in \bR^d$ and a state $\o$ on $\cAu_x$ one may 
 consider the family of states $(\olu)_{\l > 0}$ on
$\cAu_x$ defined by
\begin{equation}
\olu(\Au) := \o(\Aul)\,, \quad \l > 0\,, \ \Au \in \cAu_x\,,
\end{equation}
as a net (generalized sequence) of states indexed by the positive
reals and directed towards $\l = 0$. 
 This net of states on the $C^*$-algebra $\cAu_x$
possesses weak-* limit points as $\l \to 0$. The collection
of these limit points is denoted by $SL_x(\o) = \{\ooi,\, \iota \in \bI_x\}$
where $\bI_x$ is some suitable index set labelling the collection of
limit points.
The states in $SL_x(\o)$ are called scaling limit states of $\o$ at
$x$. Proceeding as in \cite{BV1} one now forms the
GNS-representation $(\poi,\Hoi,\Ooi)$ of $\cAu_x$ corresponding to an
$\ooi \in SL_x(\o)$. It induces a net of $C^*$-algebras
\begin{equation}
 \cO \to \Aoi(\cO) := \poi(\cAu_x(\cO))\,,
\end{equation}
called the scaling limit net of the scaling limit state $\ooi$, and
provided that ${\rm ker}\,\poi$ is left invariant under the action of
the lifted translations $\{\ayu\}_{y \in \bR^d}$, there is an induced action
\begin{eqnarray}
 \ayoi(\poi(\Au))&: = &\poi(\ayu(\Au)) \,,\\
 \ayoi(\Aoi(\cO)) & = & \Aoi(\cO + y)\,, \quad y \in \bR^d\,, \ \Au \in
 \cAu_x\,,
\end{eqnarray}
of the translations by strongly continuous $C^*$-automorphisms on that
scaling limit net.

Recall that a state $\o$ of the underlying theory is called a {\it
  vacuum state} with respect to the translation group $\{\a_y\}_{y \in
  \bR^d}$ if for all $A,B \in \cA$ the support of the
Fourier-transform of $y \mapsto \o(A^*\a_y(B))$ lies in the forward
lightcone $\overline{V}_+$. This implies that $\o$ is translationally
invariant as a consequence of the following standard argument: Since
$\o$ is a positive functional, it is hermitean, and so the stated
constraint on the Fourier-spectrum of the action of the translations
entails that the Fourier-transform of the bounded function
 $x \mapsto \o(\a_x(A))$ has, for
all $A = A^* \in \cA$, just the origin as its support. Hence it
follows that the function $x \mapsto \o(\a_x(A))$ must be constant,
and by linearity, this extends to arbitrary $A \in \cA$. Considering
the  GNS-representation $(\pi,\cH,\O)$ of $\cA$ corresponding to
$\o$, the theory $(\cO \to \pi(\cA(\cO)),\{\a^{\p}_x\}_{x \in \bR^d})$
is then a theory fulfilling $(SC)$, where $\a^{\p}_x \lcrc \p = \p
\lcrc \a_x$ is the induced action of the translations. Moreover, $\O$
is a translation-invariant vacuum vector. Conversely, a theory
fulfilling $(SC)$ and possessing an invariant vacuum vector $\O$
has a vacuum state $\o(\,.\,) = \langle \O,\,.\,\O\rangle$.

It is easily proved that, if the
underlying theory fulfills the locality condition $(L)$, then the
scaling limit nets $\cO \to \Aoi(\cO)$ corresponding to all $\ooi \in
SL(\o)$ for any state $\o$ on $\cA$ fulfill locality as well.
Furthermore,
if the underlying theory admits a vacuum state $\o$, then one can show
that each scaling limit state $\ooi \in SL_x(\o)$ is a vacuum state on
$\cAu_x$ with respect to the lifted translations $\{\ayu\}_{y \in
  \bR^d}$ \cite{BV1}. 
At the present level of generality, where we don't assume that the
underlying theory possesses a vacuum state, we don't know if any of
the scaling limit states are vacuum states on the scaling limit
algebra $\cAu_x$. But it turns out that certain constraints on
$ACS^2(\o)$ for a state $\o$ of the underlying theory suffice to
conclude that its scaling limit states are vacuum states. More
precisely, we obtain the following statement.
\begin{Thm}
Let $x \in \bR^d$, $\xx = (x,x) \in (\bR^d)^2$, and let $\o$ be any
state of the underlying theory (i.e.\ $\o$ is a positive,
normalized functional on $\cA$).
\\[6pt]
(a) \quad Suppose that $ACS_{\xs}^2(\o) \subset \cV_2 \backslash
\{0\}$. Then each scaling limit state $\ooi \in SL_x(\o)$ is a
  translationally invariant vacuum state on $\cAu_x$.
\\[6pt]
(b) \quad If the underlying theory statisfies the condition of
locality and if  $ACS_{\xs}^2(\o) = \emptyset$, then for each $\ooi \in
SL_x(\o)$ the scaling limit algebras $\Aoi =
\overline{\bigcup_{\cO}\Aoi(\cO)}^{C^*}$ are Abelian.
\end{Thm}
\noindent
{\it Remarks.} (i) In general, the scaling limit states $\ooi$ need
not be pure states on $\cAu_x$. It is shown in \cite{BV1} that, if the
underlying theory has a pure vacuum state and fulfills locality, then
the  scaling limit states will be pure vacuum states for $d \ge 3$ (but
not for $d = 2$, cf.\ \cite{BV2,Bu3})
\\[6pt]
(ii) The situation that all scaling limit algebras are Abelian is in
\cite{BV1} referred to by saying that $\o$ has a ``classical scaling
limit'', motivated by the fact that an Abelian algebra doesn't
describe a quantum theory. In \cite{BV1} it was moreover asssumed that
$\o$ is a pure vacuum state which leads for $d \ge 3$ to the much
stronger conclusion that $\Aoi = \bC1$ for all Abelian scaling limit algebras
\cite{Bu2}.
\begin{proof}
(a) The statement is proved once we have shown that for any $f
\in \cS(\bR^d)$ whose Fourier-transform $\widehat{f}$ has compact
support in $\bR^d\backslash \overline{V}_+$ there holds 
\begin{equation}
 \olu(\Au^*\afu(\Bu)) \to 0 \quad {\rm as}\quad \l \to 0
\end{equation}
for all $\Au,\Bu \in \cAu_x^{\lcrc} = \bigcup_{\cO}\cAu_x(\cO) \subset
{\bf A}_x$ where 
\begin{equation}
  (\afu(\Bu))_{\l} := \int f(y) \a_{\l y}(\Bul)\, d^dy\,, \quad \l > 0\,.
\end{equation}
To show this, we first use the positivity of $\o$ to obtain the
estimate
\begin{eqnarray}
        |\olu(\Au^*\afu(\Bu))|^2 & \le &
        \olu(\Au^*\Au)\,\olu(\afu(\Bu)^*\afu(\Bu)) \nonumber \\
 & \le & ||\,\Au\,||^2\cdot \olu(\afu(\Bu)^*\afu(\Bu))\,.
\end{eqnarray}
Furthermore, we have for each $\l > 0$
\begin{equation}
 \olu(\afu(\Bu)^*\afu(\Bu)) = \frac{1}{\l^{2d}} \int
 \overline{f(\l^{-1}
 y)}f(\l^{-1}y')\o(\a_y(\Bul^*)\a_{y'}(\Bul))\,d^dy\,d^dy'\,. 
\end{equation}
Now let $U$ be an open neighbourhood of ${\rm supp}\,\widehat{f}$ so
that $\overline{U}$ is compact and contained in $\bR^d\backslash
\overline{V}_+$. Let $\cU := U \times - U$, then $\cU$ is an open
subset of $(\bR^d)^2\backslash\{0\}$ such that $\overline{\cU}$ is
compact and contained in $(\bR^d)^2 \backslash \cV_2$. Since
$ACS^2_{\xs}(\o) \subset \cV_2\backslash\{0\}$, one can find some
function $h \in \cD((\bR^d)^2)$ with $h(0) =1$ and the property that,
for all $\Au \otimes \Au' \in \cAu_x^{\lcrc} \otimes \cAu_x^{\lcrc}$,
\begin{equation}
 \int \eilky h(\yy) \o(\a_y(\Aul)\a_{y'}(\Aul'))\,d\yy =
 O^{\infty}(\l) \quad {\rm as} \quad \l \to 0
\end{equation}
holds uniformly for $\kk = (k,k') \in \cU$
(with the obvious notation $\yy = (y,y')$). In view of Prop.\ 3.2(a) it may
be assumed that there is a function $h_1 \in \Cin((\bR^d)^2)$ which is
supported outside a ball with some positive radius around the origin
in $(\bR^d)^2$ and such that $h + h_1 = 1$. Since $f$ is rapidly
decaying at infinity and
$\sup_{\l,y,y'}|\o(\a_y(\Bul^*)\a_{y'}(\Bul))| \le ||\,\Bu\,||^2$, one
obtains that
\begin{equation}
 \frac{1}{\l^{2d}} \int
 \overline{f(\l^{-1}
 y)}f(\l^{-1}y')h_1(\yy)\o(\a_y(\Bul^*)\a_{y'}(\Bul))\,d^dy\,d^dy' =
 O^{\infty}(\l) \quad {\rm as} \quad \l \to 0\,.
\end{equation}
Therefore, setting $\o_{\l}(\yy) := \o(\a_y(\Bul^*)\a_{y'}(\Bul))$,
the following chain of equations holds for $\l \to 0$:
\begin{eqnarray}
 \frac{1}{\l^{2d}}\int \overline{f} \otimes f (\l^{-1}\yy) \o_{\l}(\yy) \,d\yy
& = & 
\frac{1}{\l^{2d}}\int \overline{f} \otimes f (\l^{-1}\yy)
 h(\yy) \o_{\l}(\yy) \,d\yy + O^{\infty}(\l) \nonumber \\
& = & \int \overline{f} \otimes f (\yy) (h\o_{\l})(\l\yy) \,d\yy +
O^{\infty}(\l) \nonumber \\
& = & \frac{1}{\l^{2d}(2\p)^{2d}}
\int \overline{\widehat{f}(k)} \widehat{f}(-k')
\widehat{h \o_{\l}}(\l^{-1}\kk) \,d\kk + O^{\infty}(\l) \nonumber \\
& \le& \frac{1}{\l^{2d}(2\p)^{2d}} \,||\,\widehat{f} \,||_{L^1}^2\cdot 
\sup_{\kks \in \cU} \widehat{h\o_{\l}}(\l^{-1}\kk) + O^{\infty}(\l)
\nonumber \\
& = & O^{\infty}(\l)
 \end{eqnarray}
where for the last estimate we have used the bound (4.39). Comparison
with (4.37) and (4.38) shows that $\olu(\Au^*\afu(\Bu)) = O^{\infty}(\l)$ as
$\l \to 0$ for all $\Au,\Bu \in \cAu_x^{\lcrc}$ and $f \in \cS(\bR^d)$ with
$\widehat{f}$ having compact support in $\bR^d\backslash
\overline{V}_+$, which yields the result.
\\[6pt]
(b) The like argument as in (a) shows that  
$\olu(\Au^*\,\afu(\Bu)) = O^{\infty}(\l)$ as
$\l \to 0$ holds for all $\Au,\Bu \in \cAu_x^{\lcrc}$ and $f \in \cS(\bR^d)$
where the support of $\widehat{f}$ is compact and doesn't contain the
origin.
This entails that for any choice of $\Au,\Bu \in \cAu_x$ the
Fourier-transform of the bounded function $y \mapsto
\ooi(\Au^*\ayu(\Bu))$ has only the origin as its support and hence
the function is constant.  It follows that $\ayoi(B_{0,\iota}) =
B_{0,\iota}$ for all $B_{0,\iota} \in \Aoi$. Since the net $\cO \to
\Aoi(\cO)$ fulfills locality, it follows that $B_{0,\iota}$ commutes
with all elements of $\Aoi$ in view of (4.34), thus $\Aoi$ is Abelian.
\end{proof}

%%%%%%%%%%%%%%%%%%%%%%%%%%%%%%%%%%%%%%%%%%%%%%%%%%
\section{Comparison with wavefront sets of Wightman distributions}
%%%%%%%%%%%%%%%%%%%%%%%%%%%%%%%%%%%%%%%%%%%%%%%%%%
\setcounter{equation}{0}
%%%%%%%%%%%%%%%%%%%%%%%%%%%%%%%%%%%%%%%%%%%%%%%%%%
We will now specialize the setting so as to be able to compare the
asymptotic correlation spectrum with the wavefront set of Wightman
distributions. So we assume now that the local observable algebras
$\cA(\cO)$ of our given theory $(\cO \to \cA(\cO),\axR)$ are
concretely given as operator algebras on some Hilbertspace $\cH$, and
that the action of the translations is the adjoint action of a weakly
continuous unitary
group representation $\bR^d \owns x \mapsto U(x)$ on $\cH$.
To simplify the proof of Theorem 5.1 below, we will here assume that
$\cA(\cO) = \cA(\cO)''$ and relax the
condition that $\{\a_x\}_{x \in \bR^d}$ acts strongly continuously to
the requirement of weak continuity. (However, it
 can be shown that the result of Thm.\ 5.1 also obtains when
 $(\cA,\{\a_x\}_{x \in \bR^d})$ is a $C^*$-dynamical
system with the property that all operators of the form $\int
h(x)U(x)AU(x)^{-1}\,dx$ are in $\cA(\cO)$ whenever $A \in
\cA(\cO_1)''$ and $h \in \cD(\cO_2)$ with $\cO_1 + \cO_2
\subset \cO$.)

Moreover, it will be assumed that there is a Wightman quantum field $\Phi$ on
$\cH$ (cf.\ \cite{SW}),
 i.e.\ a linear map $\cD(\bR^d) \owns f \mapsto
\Phi(f)$ which assigns to each complex-valued test-function $f$ a
closable operator $\Phi(f)$ with a dense domain $D_{\Phi} \subset \cH$
which is independent of $f$ and left invariant by all the $\Phi(f)$;
additionally, it will be supposed that $\Phi(\overline{f}) \subset
\Phi(f)^*$
where $\overline{f}$ is the complex conjugate of $f$ and the star
denotes the adjoint operator.  
We also require that $f \mapsto \Phi(f)$ is an operator-valued
distribution, that is, for any $\psi,\psi' \in D_{\Phi}$, the map $f \mapsto
\langle \psi',\Phi(f)\psi\rangle$ is an element of the distribution
space $\cD'(\bR^d)$. A further assumption is the covariance of
the quantum field with respect to the translations of the given
theory, i.e.\
\begin{eqnarray}
 \ax(\Phi(f)) &=& \Phi(\t_x(f))\,, \quad x \in \bR^d,\ f \in
 \cD(\bR^d)\,, \\
 U(x)D_{\Phi}& \subset& D_{\Phi}\,, \quad x \in \bR^d\,.
\end{eqnarray}
Finally we assume that the quantum field is affiliated to the local von
Neumann algebras of the given theory. This means that,
if $I_f |\Phi(f)|$ denotes the polar decomposition of the closed
extension of $\Phi(f)$, then
$I_f$ and the spectral projections of $|\Phi(f)|$ are contained in
$\cA(\cO)''$ whenever ${\rm supp}\,f \subset \cO$.
  (Note that presently we make no
assumptions regarding locality,
spectrum condition or the existence of a vacuum state.)

The assumptions imply that for each $\psi,\psi' \in D_{\Phi}$ the ``$n$-point
functionals'' 
\begin{equation}
\varphi_n(f_1 \otimes \cdots \otimes f_n) := \langle\psi',\Phi(f_1)
\cdots \Phi(f_n) \psi \rangle\,, \quad f_j \in \cD(\bR^d)\,,
\end{equation}
define distributions in $\cD'(\bR^{dn})$. On the other hand, the
two vectors $\psi,\psi'\in D_{\Phi}$ give rise to a continuous linear
functional
\begin{equation}
 \varphi(A) := \langle\psi',A \psi \rangle \,, \quad A \in \cA\,,
\end{equation}
on the quasilocal algebra $\cA$. With this notation, the following
holds:
\begin{Thm}
$WF(\varphi_n) \subset ACS^n(\varphi)$ for all $n \in \bN$.
\end{Thm}
\begin{proof}
 For all real-valued
test-functions $f$  and $t > 0$, the operator $(1 +
t\Phi(f)^2)^{-1}\Phi(f)$ is  bounded and it holds that
\begin{equation}
 ||\, (1 + t\Phi(f)^2)^{-1}\Phi(f)\,|| \le t^{-1}\,, \quad 0 < t \le
 1\,.
\end{equation}
Here and in the following, $\Phi(f)^2$ is notationally identified with
its Friedrich's extension.
Furthermore, by the mean value theorem we have for any real $f \in
\cD(\bR^d)$, $y \in \bR^d$ and $\psi \in D_{\Phi}$,
\begin{equation}
||\,\a_y\left((1 + t\Phi(f)^2)^{-1}\Phi(f) - \Phi(f)\right)\psi\,||
\le ||\,\Phi(\t_yf)^3\psi\,||\cdot t\,,\quad t > 0 \,.
\end{equation}

Now let $n \in \bN$, and let $f_1,\ldots,f_n \in \cD(\bR^d)$
be an $n$-tupel of real-valued test-functions, and $q_1,\ldots,q_n$ an
$n$-tupel of real numbers with values not less than 1. Then we
write for $j = 1,\ldots,n$:
\begin{eqnarray}
f^{(j)}_{\l}(y) & := &f^{(j)}(\frac{y}{\l^{q_j}})\,, \quad \l > 0\,,
\ y \in \bR^d\,,\\
 S_{p_j}(\l) \equiv S^{(j)}_{p_j}(\l) &:=& (1 +
 \l^{p_j}\Phi(f_{\l}^{(j)})^2)^{-1}\Phi(f_{\l}^{(j)})\,, \quad \l >
 0\,, p_j \ge 1\,.
\end{eqnarray}
Here we appoint the convenient convention to use the index $j$ of
$p_j$ to distinguish the different $S^{(j)}$, so that the superscript
$j$ on $S^{(j)}_{p_j}$ may be dropped without losing information.

The main step in the proof of our Theorem
is to establish the following auxiliary result.
\begin{Lemma}
 Let $n \in \bN$ and suppose that the $f_j$ and $q_j$, $j =
 1,\ldots,n$ are given
 arbitrarily. Then for each $M > 0$, each compact subset $K \subset
 \bR^{dn}$ and each $\psi \in D_{\Phi}$
 one can determine numbers $p_j \ge 1$, $j = 1,\ldots,n$,
such that
$$
||\left( \a_{y_n}(S_{p_n}(\l)) \cdots
  \a_{y_1}(S_{p_1}(\l)) - \Phi(\t_{y_n}f^{(n)}_{\l}) \cdots
  \Phi(\t_{y_1}f^{(1)}_{\l}) 
\right) \psi \,|| = O(\l^M) \ \ {\rm as}\ \ \l \to 0\,,
$$
uniformly for $(y_1,\ldots,y_n) \in K$.
\end{Lemma}
\begin{proof} 
This Lemma will be proved via induction on $n$, so we begin by
demonstrating the statement for the case $n =1$.

According to the estimate (5.9), we have 
\begin{equation}
||\, \a_{y_1}\!\!\left(S_{p_1}(\l) -
  \Phi(f_{\l}^{(1)})\right)\psi\,|| \le
||\,\Phi(\t_{y_1}f^{(1)}_{\l})^3\psi\,|| \cdot \l^{p_1}\,.
\end{equation}
Now we make use of the fact that the $\varphi_n$ as in (5.3) are
distributions. Hence there is for the chosen $f_1 \in \cD(\bR^d)$
and $q_1 \ge 1$, and for any compact subset $K$ of $\bR^d$, some
number $m = m(f_1,q_1,K) \ge 0$ so that
\begin{equation}
\sup_{y_1 \in K}\, ||\,\Phi(\t_{y_1}f^{(1)}_{\l})^3\psi \,|| \le C
\cdot \l^{-m}\,, \quad 0 < \l \le 1
\end{equation}
with some $C > 0$.
Thus, when we choose for given $M > 0$ any $p_1 \ge M + m$, we obtain
\begin{equation}
\sup_{y_1 \in K}\,||\, \a_{y_1}\!\!\left(S_{p_1}(\l) -
  \Phi(f_{\l}^{(1)})\right)\psi\,|| \le C \cdot \l^M\,, \quad 0 < \l
\le 1\,,
\end{equation}
which proves the required statement for $n = 1$.

To complete the proof of the Lemma by induction, we suppose now that it
holds for some arbitrary fixed $n\in\bN$. We need to show that then it
holds also for the next integer $n + 1$. So let a set $f_j$ of
real-valued test-functions and numbers $q_j \ge 1$, $j =
1,\ldots,n+1$, be given, as well as an arbitrary $M > 0$. We introduce
the following abbreviations: $\yy = (y_1,\ldots,y_n)$, $p =
(p_1,\ldots,p_n)$,
\begin{eqnarray}
X_{p,\ys}(\l) & : = & \a_{y_n}(S_{p_n}(\l)) \cdots
  \a_{y_1}(S_{p_1}(\l))\,, \\
Y_{\ys}(\l) & : = & \Phi(\t_{y_n}f^{(n)}_{\l}) \cdots
  \Phi(\t_{y_1}f^{(1)}_{\l}) \,.
\end{eqnarray}
Then it holds that
\begin{eqnarray}
\lefteqn{ \hspace*{-1.3cm}
 || \left(\a_{y_{n+1}}(S_{p_{n+1}}(\l)) X_{p,\ys}(\l) -
   \Phi(\t_{y_{n+1}}f^{(n+1)}_{\l})Y_{\ys}(\l)\right)\psi\,||} \nonumber\\ 
 & \le & ||\,\a_{y_{n+1}}(S_{p_{n+1}}(\l)) (X_{p,\ys}(\l)
         -Y_{\ys}(\l)) \psi \,|| \\
 & & +\ ||\, \a_{y_{n+1}}(S_{p_{n+1}}(\l) -
 \Phi(f_{\l}^{(n+1)})) Y_{\ys}(\l)\psi\,||\,. 
\end{eqnarray}
One can now apply the same argument as given for the case $n = 1$ to
gain for the term (5.18) an estimate of the type
(5.12), namely 
\begin{eqnarray}
\lefteqn{ ||\,\a_{y_{n+1}}\!\!\left( S_{p_{n+1}}(\l) -\Phi(f^{(n+1)}_{\l})
  \right)Y_{\ys}(\l)\psi \,||} \\
 & \le & ||\,\Phi(\t_{y_{n+1}}f_{\l}^{(n+1)})^3Y_{\ys}(\l)\psi\,||
 \cdot \l^{p_{n+1}}\,. \nonumber
\end{eqnarray}
Again along the lines of the arguments given for the case $n =1$, we
use that the quantum field is an operator-valued distribution,
implying that for the given $f_j$ and $q_j$, $j = 1,\ldots,n+1$, and
for any compact subset $K'$ of $\bR^{d(n+1)}$, there is some number $m'
\ge 0$ (depending on the said data) so that
\begin{equation}
 \sup_{(\ys,y_{n+1}) \in K'}\,
 ||\,\Phi(\t_{y_{n+1}}f_{\l}^{(n+1)})^3Y_{\ys}(\l)\psi \,|| \le
 C'\cdot \l^{-m'}\,, \quad 0 < \l \le 1\,,
\end{equation}
holds with a suitable $C' > 0$. So choosing for given $M' > 0$ a
$p_{n+1} \ge M' + m'$, it follows that
\begin{equation}
\sup_{(\ys,y_{n+1}) \in K'}\,
 ||\,\a_{y_{n+1}}\!\!\left( S_{p_{n+1}}(\l) -\Phi(f^{(n+1)}_{\l})
  \right)Y_{\ys}(\l)\psi \,|| \le
 C'\cdot \l^{M'}\,, \quad 0 < \l \le 1\,.
\end{equation} 
With this choice of $p_{n+1}$ one gets, in view of (5.5),
\begin{eqnarray}
\lefteqn{ \hspace*{-3.0cm}
 ||\,\a_{y_{n+1}}(S_{p_{n+1}}(\l))(X_{p,\ys}(\l) -
 Y_{\ys}(\l))\psi\,|| \le ||\,S_{p_{n+1}}(\l)\,||\,||\,(X_{p,\ys}(\l)
 -Y_{\ys}(\l))\psi\,|| \nonumber }\\ 
& \le & \l^{-p_{n+1}}\cdot||\,(X_{p,\ys}(\l)
 -Y_{\ys}(\l))\psi\,||\,, \quad 0 < \l \le 1\,.
\end{eqnarray}
However, by the induction hypothesis, the statement of the Lemma holds
for the fixed $n$, and so we may conclude that for the given
$f_j,q_j$, $j = 1,\ldots,n$, and $M = M' + p_{n+1}$ we find numbers
$p_1,\ldots,p_n \ge 1$ with the property
\begin{equation}
\sup_{\ys \in K^{\lcrc}} \, ||\, (X_{p,\ys}(\l) - Y_{\ys}(\l))\psi\,||
= O(\l^{M})
\end{equation}
where $K^{\lcrc}$ denotes the projection of $K'$ onto the first $n$
entries of vectors in $\bR^d$. Combining this with (5.14-19),
 we see the induction hypothesis that the statement of the Lemma
holds for some arbitrary $n \in \bN$ to imply the validity of the
statement for the subsequent integer $n+1$. This proves the Lemma.
\end{proof} % Lemma
Continuing the proof of the theorem, our task is now to show that
$(\xx;\ks) \nin ACS^n(\varphi)$ implies $(\xx;\ks) \nin
WF(\varphi_n)$. So suppose $(\xx;\ks) \nin ACS^n(\varphi)$. Now let $f
\in \cD(\bR^d)$ with $\widehat{f}(0)= 1$ and let, for an arbitrarily
given set of numbers $q_1,\ldots,q_n \ge 1$,
\begin{equation}
 f_{\l}^{(j)}(x') := \theta(\l)f(\frac{x' - x_j}{\l^{q_j}})\,, \quad
 x' \in \bR^d\,,\ \l > 0\,,
\end{equation}
where $\theta(\l)$ is a cut-off function, $\theta(\l) = 1$ for $0<
\l<1$, $\theta(\l) = 0$ for $\l \ge 1$. Then let $p_j \ge 1$, $j =
1,\ldots,n$, and let $S_{p_j}(\l)$ be defined as in (5.8) with the
$f_{\l}^{(j)}$ of (5.21). It is now easily seen that each testing
family $\lb A_{\l}^{(j)}\rb$, $j = 1,\ldots,n$, defined by 
$A^{(j)}_{\l}= S_{p_j}(\l)$, is an element of ${\bf A}_{\xs}$.
  We conclude that 
there are an open neighbourhood $V$ of $\ks$ and $h \in \cD(\bR^{dn})$
with $h(0) = 1$ such that 
\begin{equation}
 \int \eilky h(\yy)\varphi(\a_{y_1}(S_{p_1}(\l)) \cdots
 \a_{y_n}(S_{p_n}(\l))) \, d\yy = O^{\infty}(\l) \quad {\rm as} \quad
 \l \to 0
\end{equation}
uniformly in $\kk \in V$. Since the $p_1,\ldots,p_n \ge 1$ are
arbitrary, application of Lemma 5.2 yields that this last relation
entails 
\begin{equation}
\int \eilky h(\yy) \varphi_n(\t_{y_1}f_{\l}^{(1)} \otimes \cdots
\otimes \t_{y_n}f_{\l}^{(n)}) \,d\yy
 = O^{\infty}(\l) \quad {\rm as} \quad \l \to 0
\end{equation}
uniformly in $\kk \in V$, where the $f_{\l}^{(j)}$ are of the form
(5.21) with $q_1,\ldots,q_n \ge 1$ given arbirarily. Therefore,
comparison with the part (a) $\Leftrightarrow$ (c) of Prop.\ 2.1 shows
that (5.23) just expresses that $(\xx;\ks) \nin WF(\varphi_n)$.
\end{proof} % Theorem

Finally we present a statement guaranteeing that quite generally the
wavefront set of the $2n$-point distributions
$\varphi_{2n}(f_1,\ldots,f_{2n}) = \langle
\psi,\Phi(f_1),\ldots,\Phi(f_{2n})\psi\rangle$ for $\psi$ in the domain
of the Wightman field $\Phi$ are non-empty. Following are our
assumptions: We consider a theory $(\cO \to \cA(\cO),\{\a_x\}_{x \in
  \bR^d})$ given in a concrete Hilbertspace representation on a
Hilbertspace $\cH$ together with a quantum field $\Phi$ satisfying the
assumptions listed at the beginning of the section (so that $\Phi$ is
affiliated to the local von Neumann algebras). Additionally, we
suppose that the theory fulfills locality $(L)$, spectrum condition
$(SC)$ and also that there exists an up to a phase unique vacuum
vector $\O \in \cH$ which is cyclic for the algebra $\cA$.
 If $\cO \subset \bR^d$ is any open neighbourhood
of the origin in $\bR^d$, we denote by $\cO_x := \cO + x$  the
$\cO$-neighbourhood of $x$. With these conventions, we get:
\begin{Prop}
Let $n \in \bN$ and $\xx =(x_1,\ldots,x_n) \in \bR^{dn}$, and let
$\psi \in D_{\Phi}$ be a unit vector which is separating
for the local von Neumann algebras of the theory. Let $\cO$ be some
neighbourhood of the origin in $\bR^d$ and suppose that for each
choice of $f_j \in \cD(\cO_{x_j})$, $j = 1,\ldots,n-1$, there is some
continuous function $K_{f_1,\ldots,f_{n-1}}$ on $\cO_{x_n} \times
\cO_{x_n}$ with values in $\bC$ such that
\begin{equation}
\langle\psi,\Phi(\overline{f_1})\cdots\Phi(\overline{f_{n-1}})\Phi(g)
\Phi(h)\Phi(f_{n-1})\cdots\Phi(f_1)\psi\rangle
 = \int K_{f_1,\ldots,f_n}(y,y')g(y)h(y')\,d^dyd^dy'
\end{equation}
holds for all $g,h \in \cD(\cO_{x_n})$.

 Then it follows that the field
operators are multiples of $1$, i.e.\ for each $f \in \cD(\bR^d)$
there is some $c_f \in \bC$ such that $\Phi(f) = c_f1$.
\end{Prop}
\begin{proof}
To simplify notation we will assume that $x_n = 0$. Let $\cO'$ be
another neighbourhood of the origin in $\bR^d$ so that
$\overline{\cO'} \subset \cO$. Then one can determine some real-valued
$g \in \cD(\bR^d)$ so that the sequence $g_{\n}$ with $g_{\n}(y) =
{\n}^dg(\n y)$, $\n \in \bN$, converges for $t\to 0$ to the $\d$-distribution
concentrated at the origin and has, moreover, the property that
$g^{(x')}_{\n} := \t_{x'}(g_{\n}) \in \cD(\cO)$ for all $x' \in
\cO'$. For each choice of $x' \in \cO'$ and
 $f_j \in \cD(\cO_{x_j})$, $j = 1,\ldots,n$,
one can then see that the sequence of vectors
\begin{equation}
 \psi_{\n}^{(x')} := (i + \Phi(g^{(x')}_{\n}))\Phi(f_{n-1})\cdots
 \Phi(f_1)\psi \,, \quad \n \in \bN\,,
\end{equation}
converges strongly as $\n \to \infty$ to some vector which we call
$\psi^{(x')}_{\infty}$. 

Consider, on the other hand, the sequence of resolvent operators
$R^{(x')}_{\n} := (i + \Phi(g^{(x')}_{\n}))^{-1}$. Notice that we have
\begin{equation}
R_{\n}^{(x')} = U(x')R^{(0)}_{\n}U(-x')\,;
\end{equation}
moreover, it holds that $R_{\n}^{(0)} \in
\cA(\frac{1}{\n}\cO^{\lcrc})''$ for some neighbourhood $\cO^{\lcrc}$ of
the origin in $\bR^d$. Since from our assumptions it follows that
$\bigcap_{\n \in \bN} \cA(\frac{1}{\n}\cO^{\lcrc})'' = \bC1$
\cite{Wigh}, an argument due to Roberts \cite{Rob} shows that
$R^{(x')}_{\n}$, $\n \in \bN$, possesses a subsequence which converges
weakly to a multiple $c_{x'}1$ of the unit operator for some $c_{x'}
\in \bC$. To ease notation, we identify $R^{(x')}_{\n}$, $\n \in \bN$,
 with this
subsequence. In view of (5.26) one deduces that $c_{x'} \equiv c$ is
independent of $x' \in \cO'$. Therefore, we obtain for each $x' \in
\cO'$, each $\psi' \in \cH$ and each choice of $f_j \in
\cD(\cO_{x_j})$, $j = 1,\ldots,n-1$:
\begin{equation}
\langle\psi',\Phi(f_{n-1})\cdots\Phi(f_1)\psi\rangle = \langle
\psi',R_{\n}^{(x')}\psi^{(x')}_{\n}\rangle 
  \to c\langle \psi',\psi^{(x')}_{\infty}\rangle \quad {\rm as} \quad
 \n \to \infty\,.
\end{equation}
Let us now distinguish two possibilities: $c = 0$ and $c \ne 0$.
\\[6pt]
$c = 0$: This possibility will again be subdivided according to the
subsequent two cases:
\\[6pt]
$n = 1$: In that case (5.27) modifies to
\begin{equation}
\langle\psi',\psi\rangle =  c\langle\psi',\psi^{(x')}_{\infty}\rangle = 0,
\quad \psi'\in \cH\,,
\end{equation}
and hence $\psi = 0$. But this is impossible since $\psi$ was required
to be separating for the local von Neumann algebras. Thus $c=0$ is
excluded for $n=1$.
\\[6pt]
$n \ge 2$: For this case we obtain
\begin{equation}
 \langle \psi',\Phi(f_{n-1})\cdots\Phi(f_1)\psi\rangle = 0\,, \quad
 \psi' \in \cH\,,
\end{equation}
for all $f_j \in \cD(\cO_{x_j})$, $j = 1,\ldots,n-1$. Since $\psi$ is
separating for the local von Neumann algebras to which the field operators
are affiliated, this last equation entails that 
\begin{equation}
 \Phi(f_{n-1})\cdots\Phi(f_1) =0
\end{equation}
holds for all real-valued $f_j \in \cD(\cO_{x_j})$.
To see this, note first that $B := \Phi(f_{n-1})\cdots\Phi(f_1)$ is
closable because it has a densely defined adjoint. Then a variation of
the argument leading to Lemma 5.2 shows that $S(t) = (1 +
t^{p_{n-1}}|\Phi(f_{n-1})|^2)^{-1}\Phi(f_{n-1})\cdots
(1 + t^{p_{1}}|\Phi(f_{1})|^2)^{-1}\Phi(f_{1})$, $t > 0$, converges
for $t \to 0$
strongly on $D_{\Phi}$ to $B$ upon suitable choice of numbers
$p_1,\ldots,p_{n-1} \ge 1$. Since $S(t) \in \cA(\tilde{\cO})''$, $t >
0$, for some large enough region $\tilde{\cO}$, it follows that for
any $A \in \cA(\tilde{\cO})'$ one has
\begin{eqnarray}
\lefteqn{\langle B^*\phi',A\psi''\rangle} \nonumber \\
& = & \lim_{t \to 0}\,\langle S(t)^*\psi',A\psi''\rangle \ = \ 
      \lim_{t \to 0}\,\langle \psi',S(t)A\psi''\rangle \nonumber \\
& = & \lim_{t \to 0}\,\langle \psi',AS(t)\psi''\rangle \ = \
      \langle A^*\psi',B\psi''\rangle
\end{eqnarray}
for all $\psi',\psi'' \in D_{\Phi}$ and thus $B$ is affiliated to
$\cA(\tilde{\cO})''$, cf.\ \cite[Lemma 2.3]{DSW} and also references
cited there. Now $B\psi = 0$ implies $E|B|^2\psi = 0$ for any spectral
projection of $|B|$, entailing $E|B|^2 =0$ since $\psi$ is separating for
$\cA(\tilde{\cO})$. Hence $B=0$.
 
 Eqn.\ (5.30) implies that we can find
some neighbourhood $\cO^*$ of the origin in $\bR^d$ with the
property that 
\begin{equation}
\Phi(\t_{x_{n-1} + a_{n-1}}(f_{n-1})) \cdots \Phi(\t_{x_1 + a_1}(f_1))
 = 0
\end{equation}
holds for all $f_j \in \cD(\cO^*)$ and $a_j \in \cO^*$, $j =
1,\ldots,n-1$.
Consequently, for each $f \in \cD(\cO^*)$ and all $b_1,\ldots,b_{n-1}
\in \cO^*_1$, a sufficiently small neighbourhood of $0 \in \bR^d$,
there holds
\begin{equation}
  U(x_{n-1} +
 b_{n-1})\Phi(f)U(x_{n-2}-x_{n-1} + b_{n-2})\Phi(f) 
 \cdots U(x_1-x_2 + b_1)\Phi(f)\O = 0\,;
\end{equation}
here the $b_j$ are the difference variables $b_{n-1} = a_{n-1}$,
$b_{n-2} = a_{n-2} - a_{n-1}$,\ldots,$b_1 = a_1 -a_2$.
Due to the spectrum condition $(SC)$, the expression on the left hand
side of (5.33) is, with respect to the variables $b_1,\ldots,b_{n-1}$,
 the boundary value of a function which is 
analytic in the tube $(\bR^d + i V_+)^{n-1}$ so that, as a consequence
of the right hand side of (5.33), it must in fact vanish for all $b_j
\in \bR^d$, $j = 1,\ldots,n-2$. It follows that $\Phi(f)^{n-1}\O = 0$,
$f \in \cD(\cO^*_1)$
and, since the vacuum vector $\O$ is separating for the local von
Neumann algebras, we conclude that  $\Phi(f) = 0$ 
 for all $f \in \cD(\cO^*_1)$. By covariance under
translations and linearity of the field operators in the
test-functions, we see that $\Phi(f) = 0$ holds for all $f \in
\cD(\bR^d)$. This verifies the statement of our Proposition in the
case $c = 0$, $n \ge 2$.
\\[6pt]
$c \ne 0$. In view of (5.27), this entails that
\begin{equation}
\psi^{(x')}_{\infty} = \lim_{\nu \to \infty}\,
 \Phi(g_{\n}^{(x')})\Phi(f_{n-1})\cdots\Phi(f_1)\psi =
 c'\Phi(f_{n-1})\cdots \Phi(f_1) \psi
\end{equation}
holds for all $f_j \in \cD(\cO_{x_j})$, $j = 1,\ldots,n-1$, and $x' \in
\cO'$, where $c' = \frac{1- i c}{c}$. Since $\lim_{\n \to \infty}\,
\int dx'\,h(x')\Phi(g_{\n}^{(x')}) = \Phi(h)$ holds weakly on
$D_{\Phi}$, for any $h \in \cD(\bR^d)$, we obtain from (5.34) the relation
\begin{equation}
(\Phi(h) - c'_h1)\Phi(f_{n-1})\cdots \Phi(f_1)\psi = 0
\end{equation}
for all $f_j \in \cD(\cO_{x_j})$, $j = 1,\ldots,n-1$, and $h \in
\cD(\cO')$, with $c'_h = c'\int dx'\,h(x')$.
As $\psi$ is by assumption separating for the local von Neumann
algebras, it follows for real-valued test-functions that
\begin{equation}
(\Phi(h) - c'_h1)\Phi(f_{n-1})\cdots \Phi(f_1) = 0\,, \quad  h \in
\cD(\cO')\,,\
f_j \in \cD(\cO_{x_j}),\  j = 1,\ldots,n-1 \,. 
\end{equation}
In the case $n=1$ we have immediately $\Phi(h) = c'_h1$. Otherwise
 we conclude as before that there is a neighbourhood $\cO^*$ of the
origin in $\bR^d$ so that
\begin{equation}
(\Phi(\t_{a_n}h) - c'1)\Phi(\t_{x_{n-1} +
  a_{n-1}}(h))\cdots\Phi(\t_{x_1 + a_1}(h)) = 0\,, \quad
a_1,\ldots , a_n \in \cO^*,\ h \in \cD(\cO^*)\,.
\end{equation}
By the same analyticity argument, based on the spectrum condition
$(SC)$, which we just used in the above case, it follows that the last
relation actually holds for all $a_1,\ldots,a_n \in \bR^d$. As a
consequence, we obtain that 
\begin{equation}
 |\Phi(\t_a h) - c'_h1|^2|\Phi(\overline{h})|^{2(n-1)} = 0
\end{equation}
for all $h \in \cD(\cO^*)$ and all $a \in \bR^d$.
Now we evaluate this relation on the vacuum state and get for all $h
\in \cD(\cO^*)$
\begin{equation}
0 = \langle\O,|\Phi(\t_ah) - c'_h1|^2|\Phi(\overline{h})|^{2(n-1)}\O\rangle \to
||\,(\Phi(h) - c'1)\O\,||\,||\,\Phi(\overline{h})^{n-1}\O\,|| 
\end{equation}
as $a$ tends to spacelike infinity because of asymptotic spacelike
clustering \cite{SW}. Since the vacuum vector $\O$ is separating for
the local von Neumann algebras we see, as before, that
$\Phi(h) = c'_h1$ or $\Phi(h) = 0$ for all $h \in \cD(\cO^*)$. Using
translation-covariance and linearity of the field operators, finally
there results $\Phi(h) = c'_h1$ or $\Phi(h) = 0$ for all $h \in
\cD(\bR^d)$, thus proving the claimed statement in the case $c \ne 0$,
and so the proof is complete.
\end{proof}
For a theory $(\cO \to \cA(\cO), \{a_x\}_{x \in \bR^d})$ with
affiliated quantum field $\Phi$ satisfying the assumptions of the last
Proposition we thus obtain:
\begin{Cor}
If the field operators $\Phi(f)$, $f \in \cD(\bR^d)$, are not all
multiples of the unit operator, then for each separating vector $\psi
\in D_{\Phi}$ and each $\xx \in (\bR^d)^n$, $n \in \bN$, the points
$(\xx,\bar{\xx}) \in (\bR^d)^{2n}$ must be contained in the singular
support of the $2n$-point distribution
\begin{equation}
\varphi_{2n}(f_1,\ldots,f_{2n}) =
\langle\psi,\Phi(f_1)\cdots\Phi(f_{2n})\psi\rangle \,, \quad
f_1,\ldots,f_{2n} \in \cD(\bR^d)\,.
\end{equation}
\end{Cor} 

%%%%%%%%%%%%%%%%%%%%%%%%%%%%%%%%%%%%%%%%%%%%%%%%%%
\section{Summary and outlook}
%%%%%%%%%%%%%%%%%%%%%%%%%%%%%%%%%%%%%%%%%%%%%%%%%%

\noindent
We have seen that it is possible to interpret the wavefront set of a
distribution as an asymptotic form of the spectrum with respect to the
translation group when the distribution is asymptotically localized at
a point. Motivated by this observation we have defined the notion of
an asymptotic correlation spectrum of states (and linear functionals)
in a generic quantum field theory in operator algebraic formulation;
this notion generalizes the wavefront set in this more general
setting. The properties of the asymptotic correlation spectrum which
we have derived support this point of view.

However, the present work investigates the asymptotic correlation
spectrum only at a preliminary stage. There are several points calling
for clarification and further development. For instance, it surely is
to be expected that the inclusion stated in Thm.\ 5.1 is proper, owing
to the fact that ${\bf A}_x$ is an algebra, so the testing families
can be multiplied and typically the spectrum is augmented under such
multiplication. To understand this relation better, an idea would be
to introduce analogues of spectral subspaces (cf.\ \cite{Arv}) in our
asymptotic correlation spectrum setting.

Since we have the hope that eventually it should be possible to use
microlocal analytic methods in the structural analysis of general
quantum field theories in curved spacetime, the next step is to
formulate an appropriate variant of the asymptotic correlation
spectrum for quantum field theory in curved spacetime in the
operator algebraic framework. Furthermore, it seems necessary to give a
formulation of the polarization set (the generalization of the
polarization set for distributions on sections in vector bundles)
\cite{Den} within the operator algebraic approach
once one attempts to formulate anything like a spin-statistics
relation in this general setting. A development in this direction will
have to address the question of how to define the concept of ``spin''
in the operator algebraic approach to quantum field theory in curved
spacetime, and its relation to spectral properties.
\\[15pt]
%%%%%%%%%%%%%%%%%%%%%%%%%%%%%%%%%%%%%%%%%%%%%%%%%%
{\Large \bf Acknowledgments}
%%%%%%%%%%%%%%%%%%%%%%%%%%%%%%%%%%%%%%%%%%%%%%%%%%
\\[8pt]
I would like to thank R.\ Brunetti, D.\ Buchholz and K.\ Fredenhagen
for interesting and instructive discussions related to the material of
this work. Moreover thanks is due to D.\ Buchholz for useful comments
on the manuscript.


\small

\begin{thebibliography}{[22]}
\bibitem{Arv} Arveson, W., ``On groups of automorphisms of operator
  algebras'', J. Funct. Anal. {\bf 15}, 217 (1974)
\vspace*{-6pt}
\bibitem{Bor2} Borchers, H.-J., ``Energy and momentum as observables
  in quantum field theory'', Commun. Math. Phys. {\bf 2}, 49 (1966)
\vspace*{-6pt}
\bibitem{Bor1} Borchers, H.-J., {\it Translation group and particle
    representations in quantum field theory}, Berlin, Springer LNP
  m40, 1996
\vspace*{-6pt}
\bibitem{BrIa} Bros, J., Iagolnitzer, D., ``Causality and local
  analyticity: a mathematical study'', Ann. Inst H. Poincar\'e {\bf A
    18}, 174 (1973)
\vspace*{-6pt}
\bibitem{BF} Brunetti, R., Fredenhagen, K., ``Interacting quantum
  fields in curved space: Renormalizability of $\phi^4$'', in: {\it
    Operator algebras and quantum field theory}, ed. S. Doplicher,
  R. Longo, J. E. Roberts, L. Zsido (Proc. of the Conf. on Operator
  Algebras, Rome, 1996), International Press, 1997
\vspace*{-6pt}
\bibitem{BFK} Brunetti, R., Fredenhagen, K.,
K\"ohler, M., ``The microlocal spectrum condition and
Wick polynomials of free fields in curved spacetimes",
Commun.\ Math.\ Phys. {\bf 180}, 633 (1996)
\vspace*{-6pt}
\bibitem{Bu1} Buchholz, D., ``On the manifestations of particles``,
  in:
{\it Mathematical physics towards the 21st century}, ed. R. N. Sen, A.
Gersten, Ben-Gurion University Press, 1994 
\vspace*{-6pt}
\bibitem{Bu2} Buchholz, D., ``Phase space properties of local
  observables and structure of scaling limits'',
  Ann. Inst. H. Poincar\'e {\bf A 64}, 433 (1996)
\vspace*{-6pt}
\bibitem{Bu3} Buchholz, D., ``Quarks, gluons, coluor: Facts or
  fiction?'', Nucl. Phys. {\bf B469}, 333 (1996)
\vspace*{-6pt}
\bibitem{BV1} Buchholz, D., Verch, R., ``Scaling algebras and
  renormalization group in algebraic quantum field theory'',
  Rev. Math. Phys. {\bf 7}, 1195 (1995)
\vspace*{-6pt}
\bibitem{BV2} Buchholz, D,. Verch, R., ``Scaling algebras and
  renormalization group in algebraic quantum field
  theory. II. Instructive examples'', {\sl hep-th/9708095}.
 To appear in Rev. Math. Phys.
\vspace*{-6pt}
\bibitem{Den} Dencker, N., ``On the propagation of polarization sets
  for systems of real principal type'', J. Funct. Anal. {\bf 46}, 351 (1982)
\vspace*{-6pt}
\bibitem{DSW} Driessler, W., Summers, S.J., Wichmann, E. H., ``On the
  connection between quantum fields and von Neumann algebras of local
  operators'', Commun. Math. Phys. {\bf 105}, 49 (1986)
\vspace*{-6pt}
\bibitem{Dui} Duistermaat, J.\ J., {\it Fourier integral operators},
  New York, Courant Institute of Mathematical Sciences, 1973
\vspace*{-6pt}
\bibitem{DH} Duistermaat, J.\ J., H{\"o}rmander, L., ``Fourier
  integral operators II'', Acta Math.\  {\bf 128}, 183 (1972)
\vspace*{-6pt}
\bibitem{Ful} Fulling, S. A., {\it Aspects of quantum field theory in
    curved space-time}, Cambridge, University Press, 1989
\vspace*{-6pt}
\bibitem{GuiSt} Guillemin, V., Sternberg, S., {\it Geometric
    asymptotics}, Providence, Amer. Math. Soc. Surveys 14, 1977
\vspace*{-6pt}
\bibitem{Haag} Haag, R., {\it Local quantum physics},
2nd ed., Berlin, Springer, 1996
\vspace*{-6pt}
\bibitem{HK} Haag, R., Kastler, D., ``An algebraic approach to quantum
  field theory'', J. Math. Phys. {\bf 5}, 848 (1964)
\vspace*{-6pt}
\bibitem{Haw} Hawking, S. W., ``The chronology protection
  conjecture'', Phys. Rev. {\bf D47}, 2388 (1993)
\vspace*{-6pt}
\bibitem{HorFI} H{\"o}rmander, L., ``Fourier integral operators I'',
  Acta Math.\ {\bf 127}, 79 (1971)
\vspace*{-6pt}
\bibitem{Hor1} H{\"o}rmander, L., {\it The analysis of linear partial
differential operators, Vol.\ 1}, Berlin, Heidelberg, New York,
Springer, 1983
\vspace*{-6pt}
\bibitem{Hor3} H{\"o}rmander, L., {\it The analysis of linear partial
differential operators, Vol.\  3}, Berlin, Heidelberg, New York,
Springer, 1985
\vspace*{-6pt}
\bibitem{Iag} Iagolnitzer, D., ``Analytic structure of distributions
  and essential support theory'', in: {\it Structural analysis of
    collision amplitudes}, ed. R. Balian,  D. Iagolnitzer,  (June
  Institute, Les Houches 1975), Amsterdam, North-Holland, 1976
\vspace*{-6pt}
\bibitem{IaSt} Iagolnitzer, D., Stapp, H. P., ``Microscopic causality
  and physical region analyticity in S-matrix theory'',
  Comm. Math. Phys. {\bf 14}, 14 (1969)
\vspace*{-6pt}
\bibitem{Jun} Junker, W., ``Hadamard states, adiabatic vacua and the
construction of physical states for scalar quantum fields on curved
spacetime'', Rev. Math. Phys. {\bf 8}, 1091 (1996)
\vspace*{-6pt}
\bibitem{KRW} Kay, B. S., Radzikowski, M. J., Wald, R. M.,
``Quantum field theory on spacetimes with a compactly generated
Cauchy-horizon'', 
Commun. Math. Phys. {\bf 183}, 533 (1997)\vspace*{-6pt}
\bibitem{Koh} K\"ohler, M., ``New examples for Wightman fields
on a manifold'', Class.\ Quantum Grav. {\bf 12}, 1413 (1995)
\vspace*{-6pt}
\bibitem{Rad1} Radzikowski, M.\ J., ``Micro-local approach to
the Hadamard condition in quantum field theory in curved space-time'',
Commun.\ Math.\ Phys.\ {\bf 179}, 529 (1996)
\vspace*{-6pt}
\bibitem{Rad2} Radzikowski, M.\ J., ``A local-to-global singularity
theorem for quantum field theory on curved space-time'',
Commun.\ Math.\ Phys.\ {\bf 180}, 1
%% -23
 (1996) 
\vspace*{-6pt}
\bibitem{Rob} Roberts, J. E., ``Some applications of dilatation
  invariance to structural questions in the theory of local
  observables'', Commun. Math. Phys. {\bf 37}, 273 (1974)
\vspace*{-6pt}
\bibitem{Sak} Sakai, S., {\it Operator algebras in dynamical systems},
Cambridge, University Press, 1991\vspace*{-6pt}
\bibitem{Sat} Sato, M., ``Hyperfunctions and partial differential
  equations'', in: {\it Proc. of the internat. conf. on functional
    analysis and related topics}, Tokyo, University 
Press, 1969.\vspace*{-6pt}
\bibitem{SW} Streater, R. F., Wightman, A. S., {\it PCT, spin and
    statistics, and all that}, New York, Benjamin, 1964 
\vspace*{-6pt}
\bibitem{Tay} Taylor, M.\ E.: {\it Pseudodifferential operators},
  Princeton, University Press, 1981
 \vspace*{-6pt}
\bibitem{Ver} Verch, R., ``Continuity of symplectically adjoint maps
  and the algebraic structure of Hadamard vacuum representations for
  quantum fields in  curved spacetime'', Rev. Math. Phys. {\bf 9}, 635
  (1997)
 \vspace*{-6pt}
\bibitem{Wald2}  Wald, R. M., {\it Quantum field theory in curved spacetime
  and black hole thermodynamics}, Chicago,  University  Press, 1994
\vspace*{-6pt}
\bibitem{Wigh} Wightman, A. S., ``La th\'eorie quantique locale et la
  th\'eorie quantique des champs'', Ann. Inst. H. Poincar\'e {\bf A
    1}, 403 (1964)
\end{thebibliography}
\end{document}